\documentclass[amsmath,amssymb,nofootinbib,notitlepage,superscriptaddress,twocolumn]{revtex4-1}
\pdfoutput=1

\usepackage{aas_macros,esint,graphicx}
\usepackage[bottom]{footmisc}
\usepackage[colorlinks=true]{hyperref}
\let\originalleft\left
\let\originalright\right
\renewcommand{\left}{\mathopen{}\mathclose\bgroup\originalleft}
\renewcommand{\right}{\aftergroup\egroup\originalright}

\newcommand{\ab}[1]{\left|#1\right|}
\newcommand{\br}[1]{\left[#1\right]}
\newcommand{\cu}[1]{\left\{#1\right\}}
\newcommand{\pa}[1]{\left(#1\right)}
\newcommand{\ed}{\mathop{}\!\mathrm{d}}
\newcommand{\pd}{\mathop{}\!\partial}
\renewcommand{\O}[1]{\mathcal{O}\pa{#1}}

\DeclareMathOperator\erf{erf}
\DeclareMathOperator\sign{sign}
\DeclareMathOperator\sn{sn}

\begin{document}

\title{Maximum Observable Blueshift from Circular Equatorial Kerr Orbiters}

\author{Delilah E.~A. Gates}
\email{dgates@g.harvard.edu}
\affiliation{Center for the Fundamental Laws of Nature, Harvard University, Cambridge, MA 02138, USA}
\author{Shahar Hadar}
\email{shaharhadar@g.harvard.edu}
\affiliation{Center for the Fundamental Laws of Nature, Harvard University, Cambridge, MA 02138, USA}
\author{Alexandru Lupsasca}
\email{lupsasca@princeton.edu}
\affiliation{Princeton Gravity Initiative, Princeton University, Princeton, NJ 08544, USA}
\affiliation{Society of Fellows, Harvard University, Cambridge, MA 02138, USA}

\begin{abstract}
The region of spacetime near the event horizon of a black hole can be viewed as a deep potential well at large gravitational redshift relative to distant observers.  However, matter orbiting in this region travels at relativistic speeds and can impart a significant Doppler shift to its electromagnetic emission, sometimes resulting in a net observed blueshift at infinity.  Thus, a black hole broadens the line emission from monochromatic sources in its vicinity into a smoothly decaying ``red wing''---whose flux vanishes at large redshift---together with a ``blue blade'' that retains finite flux up to a sharp edge corresponding to the maximum observable blueshift.  In this paper, we study the blue blade produced by isotropic monochromatic emitters on circular equatorial orbits around a Kerr black hole, and obtain simple relations describing how the maximum blueshift encodes black hole spin and inclination.  We find that small values of the maximum blueshift yield an excellent probe of inclination, while larger values provide strong constraints on spin or inclination in terms of the other.  These results bear direct relevance to ongoing and future observations aiming to infer the angular momentum of supermassive black holes from the broadening of their surrounding line emission.
\end{abstract}

\maketitle

\section{Introduction}

According to General Relativity, astrophysical black holes are entirely described by the Kerr geometry, which admits only two parameters: the mass $M$ and angular momentum $J$ of the black hole (BH).  Despite spectacular recent breakthroughs in experimental BH astrophysics \cite{LIGO2016a,EHT2019a}, it remains very challening to answer even the most elementary question about a BH:  What are its mass and angular momentum?

Several approaches to the problem of measuring BH mass have been developed.  In 2016, for instance, the Laser Interferometer Gravitational-Wave Observatory tightly constrained the masses of stellar-size BHs involved in a binary merger \cite{LIGO2016b}, establishing great promise for high-precision gravitational-wave measurements of BH parameters.  However, one cannot select specific BH targets for such measurements, which rely on the observation of rare merger events.  Traditionally, the parameters of putative BHs in the sky have been inferred from observed features in the electromagnetic spectrum of emission from their vicinity.  In the supermassive BH regime, for example, the Event Horizon Telescope recently employed Very-Long-Baseline Interferometry to measure the mass of the BH at the center of the galaxy M87 \cite{EHT2019f}, which had previously been estimated from the dynamics of nearby stars \cite{Gebhardt2011} and the surrounding gas \cite{Walsh2013}.  Earlier observations of stellar dynamics in our own galaxy also provided a precise mass measurement of the black hole Sagittarius~A* at its center \cite{Ghez2008}.

Measuring BH spin is more difficult.  The continuum-fitting method \cite{Zhang1997} has been used to measure multiple BH spins (see, e.g., Ref.~\cite{McClintock2014} and references therein), but this technique is only applicable to stellar-size BHs.  The loud emission from active galactic nuclei is widely believed to be powered by supermassive BHs, with properties of the emission (such as the presence of a narrow collimated jet) correlated with the BH spin \cite{Blandford1977}.  Thus far, the most effective approach to measuring supermassive BH spins has been the X-ray reflection method \cite{George1991}, which has already enabled over twenty such measurements (see, e.g., Refs.~\cite{Brenneman2006,Reynolds2019} and references therein), and will be the focus of this paper.  We now briefly describe this method.

Many active galactic nuclei (as well as accreting stellar-size BHs in X-ray binaries) possess a hot ``corona'' consisting of an electron-positron plasma that emits high-energy X-rays.  This X-ray emission irradiates the comparatively cool accretion flow onto the BH, exciting its upper layers and causing them to produce fluorescent emission lines at fixed X-ray frequencies, most notably the 6.4~keV FeK$\alpha$ iron line.  General relativistic effects then broaden this monochromatic ``X-ray reflection'' in a characteristic spin-dependent way.

\begin{figure*}
	\centering
	\includegraphics[width=\textwidth]{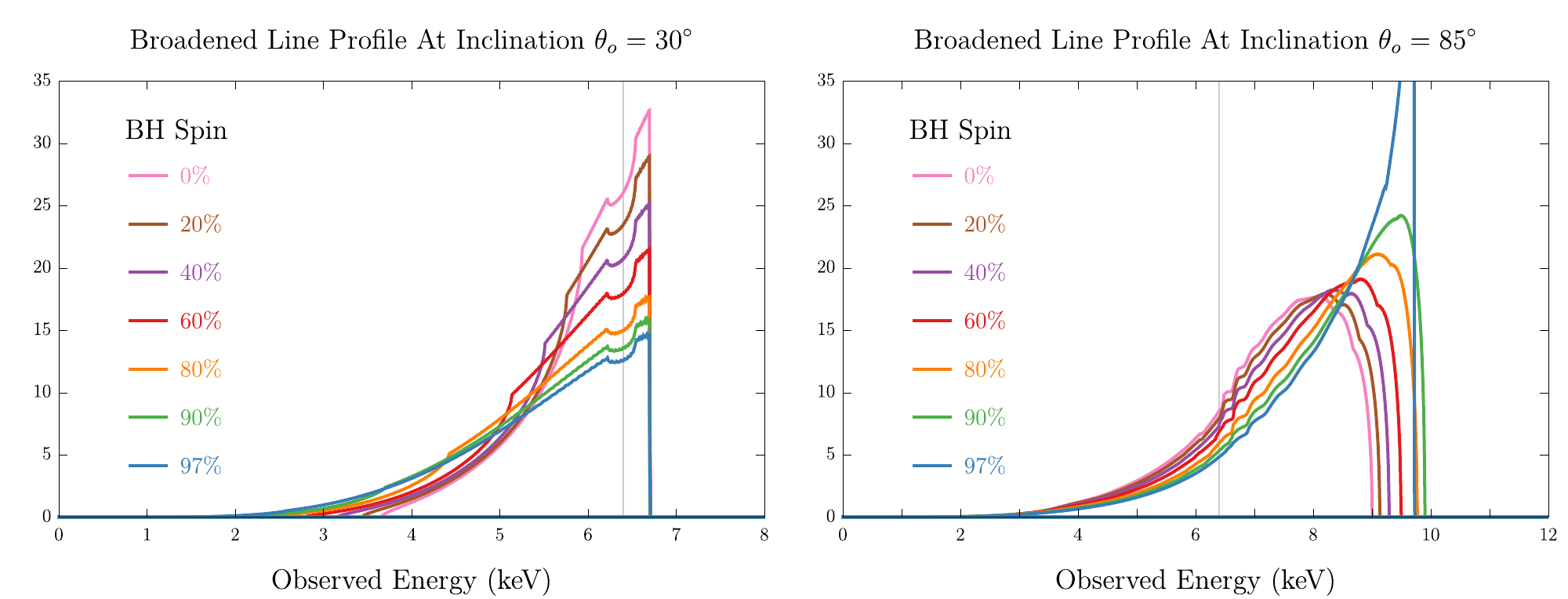}
	\caption{Specific flux density (as a function of observed photon energy, in arbitrary units) resulting from relativistic broadening of the 6.4~keV FeK$\alpha$ iron line emission from a thin accretion disk of circular equatorial orbiters.  Data simulated using the numerical code \texttt{relline} \cite{Dauser2010} courtesy of Laura Brenneman.  The space between energies at which the model is calculated is roughly equivalent to that of the future Athena/X-IFU instrument (spectral resolution $\sim$2.5~eV \cite{Barret2016}).  Prominent features of the line profile such as the red wing (long tail to the left) and blue blade (sharp edge to the right) are described in the text.}
	\label{fig:BroadenedLine}
\end{figure*}

In the idealized, analytically tractable scenario first considered by Novikov and Thorne \cite{Novikov1973}, the BH is surrounded by a geometrically thin, equatorial accretion disk of (almost) circular orbiters terminating at an inner edge coinciding with the Innermost Stable Circular Orbit (ISCO).  As the BH spin increases, the ISCO moves in closer to the horizon, and emission from the innermost part of the disk correspondingly incurs a progressively larger gravitational redshift.  The redshifted emission from these radii in the disk gives rise to an increasingly longer redshifted tail in the observed energy spectrum, with the precise fall-off of this ``red wing'' (illustrated in Fig.~\ref{fig:BroadenedLine}) depending sensitively on the BH spin.

However, gravitational redshift alone does not account for the entire morphology of the broadened line emission, which can also be Doppler shifted due to the motion of matter in the disk relative to the observer.  The Doppler blueshift imparted from radii closest to the horizon---where emitters travel at relativistic speed---may be quite large, sometimes overcoming the gravitational redshift to produce a net observed blueshift.  Thus, while the region of spacetime near the event horizon of a BH is often regarded as a deep gravity well at large redshift relative to distant observers, emission from geodesic orbiters from this region can in fact attain signficant blueshift, up to a theoretical maximum of $\sqrt{3}$ \cite{Gralla2017}.  Unlike the red wing, whose flux vanishes at high redshift, the blueshifted part of the broadened line emission retains large finite flux up to the maximum blueshift, resulting in a prominent ``blue blade'' (Fig.~\ref{fig:BroadenedLine}) that also encodes BH spin.

Applying the X-ray reflection method is challenging.  On the theory side, one must grapple with uncertainties that arise from the degeneracy of spin signatures with other physical effects, such as accretion disk winds \cite{Reynolds2019}.  On the experimental front, it has thus far proved difficult to achieve the spectral resolution required to precisely measure fine features of the broadened emission line, such as the edge of the blue blades displayed in Fig.~\ref{fig:BroadenedLine}.  Past and ongoing missions such as Suzaku, Chanda, and XMM-Newton have therefore primarily relied on observations of the redshifted tail from broadened FeK$\alpha$ line emission to estimate BH spins.  But future experiments such as the proposed Athena/X-IFU instrument will be able to measure the broadened line profile with much greater spectral resolution.  In particular, improved microcalorimetry should enable sharper measurements of the maximum observable blueshift (MOB), providing stronger constraints on BH spin.

In this paper, we investigate how BH spin and inclination are imprinted on the MOB.  To develop intuition for this relationship, we follow earlier work and study a simple tractable model consisting of a geometrically thin, equatorial disk extending down to the ISCO, and which is composed of monochromatic, isotropic emitters orbiting on timelike prograde circular equatorial geodesics.\footnote{We assume the disk is optically thin and allow photons to recross the equatorial plane on their way to the observer.  We also consider optically thick disks.  The results only differ at high spin.}  Broadened line profiles for this model may be generated using the numerical implementation \texttt{relline} \cite{Dauser2010}; these line profiles (Fig.~\ref{fig:BroadenedLine}) present several striking features.  At low inclination, the edge of the blue blade appears to be spin-independent and the MOB takes a moderate value.  At high inclination, the MOB tends to increase with spin, albeit not monotonically over the full range of spins.

These observations raise a number of questions:
\begin{enumerate}
	\item Is there a simple heuristic explanation for the common MOB at low inclination, and what is its value?
	\item What explains the qualitative shift in the MOB's spin-dependence as the inclination increases?
	\item For a fixed value of spin (inclination), what value of the inclination (spin) results in the largest MOB?
\end{enumerate}

To address these questions, we numerically compute the MOB for all spins and inclinations.  Guided by these numerics, we derive a series of complementary analytic approximations that completely characterize the MOB, providing simple answers to these questions.

At low inclination, we find that the MOB arises from light rays emitted far from the BH, where spin effects are negligible.  Taking a large emission radius as the starting point of an analytic approximation, we can then derive a simple formula for the spin-independent MOB valid at low inclination (answering Q1).  At higher inclination, the MOB originates from radii that are closer to the BH, where spin effects become relevant (answering Q2).  For equatorial observers, we prove that the MOB arises from ISCO emission and analytically derive its value.  For near-equatorial observers, our polar approximation breaks down but still helps us obtain a simple semi-analytic expression that fits to better than 1\% precision for all but near-extremal BHs.  For near-extremal BHs, we provide a formula derived from a different, earlier analytic approximation based on a high-spin expansion \cite{Lupsasca2018} (answering Q3).

Naturally, our work has some limitations.  For instance, we ignore reverberation effects \cite{Fabian2020} and leave consideration of plunging trajectories \cite{Wilkins2020} to future work.  Nonetheless, we expect the formulas that we derive and the approximations we study to provide useful heuristics that will serve as a helpful complement to more realistic numerical models.

\section{Summary of results}
\label{sec:Summary}

\begin{figure*}
	\centering
	\includegraphics[width=\textwidth]{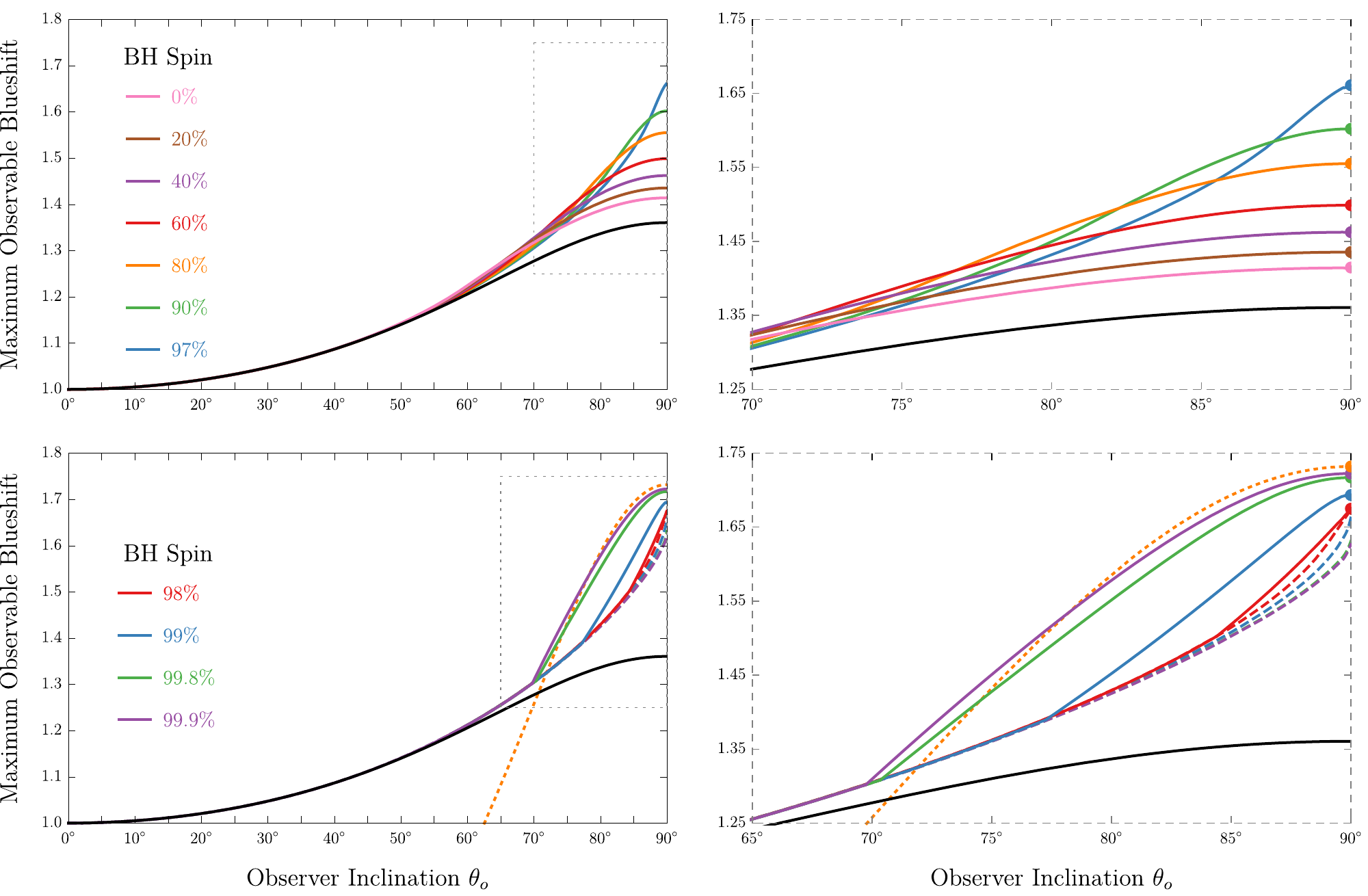}
	\caption{Maximum Observable Blueshift (MOB) from a geometrically thin, equatorial disk terminating at the Innermost Stable Circular Orbit (ISCO) of a BH, and composed of monochromatic isotropic emitters on timelike prograde circular orbits.  We numerically compute the MOB for observer inclinations $0^\circ\le\theta_o\le90^\circ$ and BH spins $0\le a<M$ (solid colored curves).  We present analytic approximations valid for low inclination and any spin [black curve, Eq.~\eqref{eq:LowInclination}], and for high inclination and high spin [orange dotted curve, Eq.~\eqref{eq:ExtremalEnvelope}].  Intermediate regimes are covered by the semi-analytic fit \eqref{eq:SemiAnalyticFit}.  The endpoint of each curve corresponds to the MOB for an equatorial observer, which we compute exactly [Eq.~\eqref{eq:ExactEquatorialMOB}].  For high spins $a/M\gtrsim97\%$, the MOB may arise from emission that encounters a large number of orbits around the BH that is logarithmically divergent in the deviation from extremality [Eq.~\eqref{eq:MaxTurningPoints}].  For the top-panel spins, the MOB always arises from emission that directly travels from the top surface of the disk to the observer.  For the bottom-panel spins, we observe at high inclinations a transition (kink in solid curves) from a regime in which the direct emission (first solid, then dashed curves) dominates the MOB to a new regime in which it is superseded by more highly blueshifted photons that encounter an additional angular turning point along their trajectory from source to observer.  This effect is only relevant for optically thin disks, while the dashed curves describe optically thick disks.  The results displayed in this figure agree to remarkable precision with the numerically simulated line profiles in Fig.~\ref{fig:BroadenedLine} above---see Fig.~\ref{fig:Comparison} below for a comparison.}
	\label{fig:MaximumBlueshift}
\end{figure*}

In this paper, we propose a target for astronomical observations to determine BH spin: the MOB of the spectal curve produced by monochromatic emitters near the BH.  Two properties of this quantity make it particularly suitable for this purpose.  First, the maximum blueshift---contrary to the redshift---is a sharp feature in the spectrum and therefore less susceptible to noise, at least for the geometrically thin accretion disks that we focus on herein.  After reviewing in Sec.~\ref{sec:Model} the parameterization $(r_s,g)$ of points in the disk by their emission radius and observed redshift, we prove in Sec.~\ref{sec:BlueBlade} that the observed flux received from a disk retains a finite nonzero value up to the MOB.  This explains the presence of the prominent ``blue blade'' feature in the spectrum.  Second, the MOB encodes information about the dimensionless BH spin $a/M=J/M^2$ and the inclination $\theta_o$ of the observer relative to the BH spin axis.  This property of the MOB is studied in the rest of the paper.  We explain our numerical procedure for computing the MOB in Sec.~\ref{sec:Numerics}, and present the derivation of our (semi-)analytic approximate results in Sec.~\ref{sec:Analytics}, relegating some technical details to App.~\ref{app:TransferFunctions}.  In the remainder of this section, we give a detailed summary of the results we obtained in each of the different regimes mentioned above.

We parameterize the observer inclination by
\begin{align}
	x=\sin{\theta_o}.
\end{align}

\subsection{Low inclination, any spin}

At low inclination $0\le x\ll1$, and for any value of spin $0\le a/M\le1$, we obtain an approximate analytic formula for the MOB [Eq.~\eqref{eq:LowInclinationMOB} below],
\begin{align}
	\label{eq:LowInclination}
	g=G(x)
	=\frac{27\sqrt{27-9x^2-6x^4}}{81\sqrt{3}-x^2\sqrt{3+2x^2}\pa{27+3x^2+2x^4}},
\end{align}
with corresponding emission radius
\begin{align}
	r_s\approx\frac{9M}{x^2\pa{1+\frac{6}{9}x^2}}.
\end{align}
In practice, this approximation is excellent for observer inclinations $0^\circ\le\theta_o\lesssim50^\circ$.  This can be be seen in Fig.~\ref{fig:MaximumBlueshift} by comparing our spin-independent formula \eqref{eq:LowInclination} (black curve) to the colored curves for the MOB at all spins.

\subsection{Equatorial observer, any spin}

In the case of an equatorial observer at $\theta_o=\pi/2$, and for any value of spin $0\le a/M\le1$, we obtain an exact analytic formula for the MOB [Eq.~\eqref{eq:EquatorialMOB} below]
\begin{subequations}
\label{eq:ExactEquatorialMOB}
\begin{align}
	g&=H(a),\qquad\Delta(r)=r^2-2Mr+a^2,\\
	H(a)&=\frac{r_\mathrm{ms}^{3/2}-2M\sqrt{r_\mathrm{ms}}+\sqrt{M}\pa{a+\sqrt{\Delta(r_\mathrm{ms})}}}{\sqrt{r_\mathrm{ms}^3-3Mr_\mathrm{ms}^2+2a\sqrt{M}r_\mathrm{ms}^{3/2}}},
\end{align}
\end{subequations}
which always arises from emission at the ISCO radius
\begin{subequations}
\label{eq:ISCO}
\begin{align}
	r_\mathrm{ms}&=M\br{3+Z_2\mp\sqrt{\pa{3-Z_1}\pa{3+Z_1+2Z_2}}},\\
	Z_1&=1+\sqrt[3]{1-a_\star^2}\pa{\sqrt[3]{1+a_\star}+\sqrt[3]{1-a_\star}\,},\\
	Z_2&=\sqrt{3a_\star^2+Z_1^2},\qquad
	a_\star=\frac{a}{M}.
\end{align}
\end{subequations}
In Fig.~\ref{fig:MaximumBlueshift} (right panels), we verify the validity of our numerics near the equator by checking that the curves end at these exact values (depicted by colored points).

\subsection{High inclination, low to moderate spin}

For high inclinations $50^\circ\lesssim\theta_o\le90^\circ$, and for any value of spin $a/M\lesssim97\%$, the approximation \eqref{eq:LowInclination} breaks down (in Fig.~\ref{fig:MaximumBlueshift}, the black curve differs from the colored curves over this range).  However, over this range of parameters, we empirically obtain the semi-analytic fit [Eq.~\eqref{eq:HighInclinationMOB} below]
\begin{subequations}
\label{eq:SemiAnalyticFit}
\begin{align}
	g&=G(x)+\br{H(a)-G(1)}\chi,\\
	\chi&=xe^{-\frac{\pa{3+7a}\pa{1-x^2}}{\sqrt{1-a^2}}},
\end{align}
\end{subequations}
which is accurate to better than 1\% precision for all inclinations and spins $a/M\le90\%$.  A slightly more complicated choice of $\chi$ [Eq.~\eqref{eq:AlternativeFit} below] holds for all $a/M\le97\%$ with precision better than 1.5\%.

\subsection{High inclination, high spin}

For high inclinations $50^\circ\lesssim\theta_o \le90^\circ$ and spins $a/M\gtrsim97\%$, the MOB no longer necessarily arises from direct emission that travels from the surface of the disk towards the observer along a modestly bent null geodesic.  Instead, the relevant photon trajectories are strongly-bent light rays that start near the event horizon and execute multiple half-orbits around the BH before escaping to infinity.  This surprising behavior was previously studied analytically using a singular high-spin expansion about extremality ($a=M$) in Refs.~\cite{Gralla2017,Lupsasca2018}.

In particular, the analytic results of Ref.~\cite{Lupsasca2018} directly give the high-spin approximation [Eq.~\eqref{eq:HighSpinMOB} below]
\begin{align}
	\label{eq:ExtremalEnvelope}
	g=\frac{\sqrt{3}x}{4-x-2x^2},
\end{align}
which is illustrated in the bottom panels of Fig.~\ref{fig:MaximumBlueshift} (orange dotted line), wherein we use dashed lines to depict the MOB from direct emission once it is superseded by a larger overall MOB from emission that circles the BH once.  In the regime $0<\kappa\ll1$ where the BH spin $a=M\sqrt{1-\kappa^2}$ approaches extremality, we find that the MOB arises from photons that encounter a (logarithmically divergent in $\kappa$) maximal number $m$ of angular turning points along their trajectory from source to observer.

Using results from Ref.~\cite{Gralla2017}, we derive an expression
\begin{align}
	\label{eq:MaxTurningPoints}
	\bar{m}^*\approx\left\lfloor\frac{1}{\pi}\log\br{\frac{18}{2+\sqrt{3}}\pa{\frac{2}{\kappa}}^{2/3}}\right\rfloor,
\end{align}
for the number of lensed disk images (which are labeled by $0\le\bar{m}\le\bar{m}^*$) relevant to the computation of the MOB at deviation from extremality $\kappa$ [Eq.~\eqref{eq:DivergentTurningPoints}].  We prove that the equatorial observer is the first to see the MOB transition from one value of $m$ to the next (in Fig.~\ref{fig:MaximumBlueshift}, new kinks in the colored curves first appear to the right and then move leftwards with increasing spin).  In practice, we numerically find that this analytical formula holds over the entire high-spin range $a/M\gtrsim97\%$.

\subsection{``Just adding one''}

It was empirically noted in Eq.~(1) of Ref.~\cite{Gralla2020a} that the arrival impact parameter $b$ of a photon emitted from an equatorial ring of Boyer-Lindquist radius $r_s$ is given by
\begin{align}
	\frac{b}{M}\approx\frac{r_s}{M}+1,
\end{align}
i.e., it is obtained by ``just adding one''.  In the course of our analysis, we derive this relation by defining a systematic expansion of $r_s$ in $M/b$, finding [Eq.~\eqref{eq:PolarTransferFunction} below]
\begin{align}
	\label{eq:TransferFunctionExpansion}
	r_s=b-M+\frac{M^2-a^2}{2b}+\frac{3M^3\pa{5\pi-16}}{4b^2}+\ldots
\end{align}
The steps behind the derivation of Eq.~\eqref{eq:TransferFunctionExpansion} are detailed in App.~\ref{app:TransferFunctions}.

\section{Model}
\label{sec:Model}

In this section, we describe in detail our disk model and the null geodesic equations that determine its optical appearance.  Most of the relevant equations were first derived by Bardeen, Press, and Teukolsky in 1972 \cite{Bardeen1972}, and by Cunningham and Bardeen in 1973 \cite{Cunningham1973}.  Our brief review follows the more recent treatment in Ref.~\cite{Gralla2017}.

\subsection{Kerr geometry and geodesics}

In Boyer-Lindquist coordinates $(t,r,\theta,\phi)$, the Kerr geometry has line element
\begin{subequations}
\begin{gather}
	ds^2=-\frac{\Delta}{\Sigma}\pa{\ed t-a\sin^2{\theta}\ed\phi}^2+\frac{\Sigma}{\Delta}\ed r^2+\Sigma\ed\theta^2\nonumber\\
	+\frac{\sin^2{\theta}}{\Sigma}\br{\pa{r^2+a^2}\ed\phi-a\ed t}^2,\\
	\Delta=r^2-2Mr+a^2,\quad
	\Sigma=r^2+a^2\cos^2{\theta}.
\end{gather}
\end{subequations}
The rotation of the Kerr BH induces a ``frame-dragging'' effect with characteristic angular frequency
\begin{align}
	\label{eq:FrameDragging}
	\omega=-\frac{g_{t\phi}}{g_{\phi\phi}}
	=\frac{2aMr}{\pa{r^2+a^2}^2-a^2\Delta(r)\sin^2{\theta}}.
\end{align}
Wordlines of constant $(r,\theta,\phi-\omega t)$ define the ``locally nonrotating frames'' (LNRFs) of the Kerr spacetime.  A particle of mass $m$ with affinely parameterized trajectory $x^\mu(\tau)$ has four-momentum $p=p_\mu\ed x^\mu$ given by
\begin{align}
	\label{eq:PhotonMomentum}
	p=-E\ed t\pm_r\frac{\sqrt{\mathcal{R}(r)}}{\Delta(r)}\ed r\pm_\theta\sqrt{\Theta(\theta)}\ed\theta+L\ed\phi,
\end{align}
given in terms of the radial and polar ``potentials''
\begin{align}
	\mathcal{R}(r)&=\br{E\pa{r^2+a^2}-aL}^2\nonumber\\
	&\qquad-\Delta(r)\br{Q+\pa{L-aE}^2+m^2r^2},\\
	\Theta(\theta)&=Q+a^2\pa{E^2-m^2}\cos^2{\theta}-L^2\cot^2{\theta},
\end{align}
which are functions of the energy $E=-p_t$, angular momentum parallel to the BH spin axis $L=p_\phi$, and Carter integral $Q=p_\theta^2+\cos^2{\theta}[a^2(m^2-p_t^2)+p_\phi^2\sin^2{\theta}]$ of the particle, all of which are conserved along the geodesic.  Photons follow null geodesics ($m=0$), which depend only on the energy-rescaled angular momentum $\lambda=L/E$ and Carter integral $\eta=Q/E^2$.

\subsection{Circular equatorial orbiters}

The Kerr geometry admits timelike circular equatorial geodesics.  At orbital radius $r=r_s$, the conserved quantities are obtained by solving $\Theta(\pi/2)=\Theta'(\pi/2)=0$ and $\mathcal{R}(r_s)=\mathcal{R}'(r_s)=0$, resulting in $Q_s=0$ and
\begin{subequations}
\label{eq:Orbiter}
\begin{align}
	\frac{E_s}{m}&=\frac{r_s^{3/2}-2M\sqrt{r_s}\pm a\sqrt{M}}{\sqrt{r_s^3-3Mr_s^2\pm2a\sqrt{M}r_s^{3/2}}},\\
	\frac{L_s}{m}&=\frac{\pm\sqrt{M}\big(r_s^2\mp2a\sqrt{Mr_s}+a^2\big)}{\sqrt{r_s^3-3Mr_s^2\pm2a\sqrt{M}r_s^{3/2}}}.
\end{align}
\end{subequations}
Hence, the four-velocity $u_s^\mu$ and angular velocity $\Omega_s$ are
\begin{align}
	\label{eq:AngularVelocity}
	u_s&=u_s^t\pa{\pd_t+\Omega_s\pd_\phi},\quad
	\Omega_s=\pm\frac{\sqrt{M}}{r_s^{3/2}\pm a\sqrt{M}},\\
	u_s^t&=\frac{r_s^{3/2}\pm a\sqrt{M}}{\sqrt{r_s^3-3Mr_s^2\pm2a\sqrt{M}r_s^{3/2}}},
\end{align}
and we see that the orbit is prograde/retrograde according to the choice of upper/lower sign in Eqs.~\eqref{eq:Orbiter}.  With respect to a LNRF, the ($\phi$-direction) orbital velocity is
\begin{align}
	\label{eq:LNRF}
	v_s=\frac{\pm\sqrt{M}\big(r_s^2\mp2a\sqrt{Mr_s}+a^2\big)}{\Delta_s^{1/2}\big(r_s^{3/2}\pm a\sqrt{M}\big)}.
\end{align}
The local rest frame of the orbiter is conveniently described using a tetrad that is aligned with $(r,\theta)$ and has the four-velocity as its time leg,
\begin{subequations}
\label{eq:OrbiterFrame}
\begin{align}
	\mathbf{e}_{(t)}&=u_s,\\
	\mathbf{e}_{(r)}&=\sqrt{1-\frac{2M}{r_s}+\frac{a^2}{r_s^2}}\pd_r,\\
	\mathbf{e}_{(\theta)}&=\frac{1}{r_s}\pd_\theta,\\
	\mathbf{e}_{(\phi)}&=v_su_s^t\pa{\pd_t+\omega_s\pd_\phi}+\sqrt{\frac{\omega_sr_s}{2aM\pa{1-v_s^2}}}\pd_\phi.
\end{align}
\end{subequations}
Everywhere along the orbit, these frame fields obey
\begin{subequations}
\begin{align}
	g_{\mu\nu}\mathbf{e}_{(a)}^\mu\mathbf{e}_{(b)}^\nu=\eta_{(a)(b)},\quad
	\eta^{(a)(b)}\mathbf{e}_{(a)}^\mu\mathbf{e}_{(b)}^\nu=g^{\mu\nu}.
\end{align}
\end{subequations}
The orbital motion is strictly stable as long as $\mathcal{R}''(r_s)<0$ and becomes marginally stable when $\mathcal{R}''(r_\mathrm{ms})=0$, with $r_\mathrm{ms}$ denoting the ISCO radius given in Eq.~\eqref{eq:ISCO} above.

The redshift of a photon emitted by an orbiter at radius $r_s$ with energy-rescaled angular momentum $\lambda$ is
\begin{align}
	\label{eq:Redshift}
	g&=\frac{E}{p^{(t)}}
	=\frac{\sqrt{r_s^3-3Mr_s^2\pm2a\sqrt{M}r_s^{3/2}}}{r_s^{3/2}\pm\sqrt{M}\pa{a-\lambda}}.
\end{align}
For fixed source radius $r_s$, the redshift $g$ is monotonically increasing/decreasing in $\lambda$ (according to whether the orbit is prograde/retrograde, respectively), since
\begin{align}
	\label{eq:RedshiftMonotonicity}
	\pd_\lambda g=\frac{\pm\sqrt{M} g}{r_s^{3/2}\pm\sqrt{M}\pa{a-\lambda}}
	\gtrless0.
\end{align}
This fact will be very useful to us, as it shows that to an observer at inclination $\theta_o$, the MOB from a source ring $r=r_s$ depends only on the maximum value of $\lambda$ attained by photons reaching the observer.  This value is determined by computing the optical appearance of the disk, a problem that we now turn to.

\subsection{Optical appearance of an equatorial disk}

\begin{figure*}
	\centering
	\includegraphics[width=\textwidth]{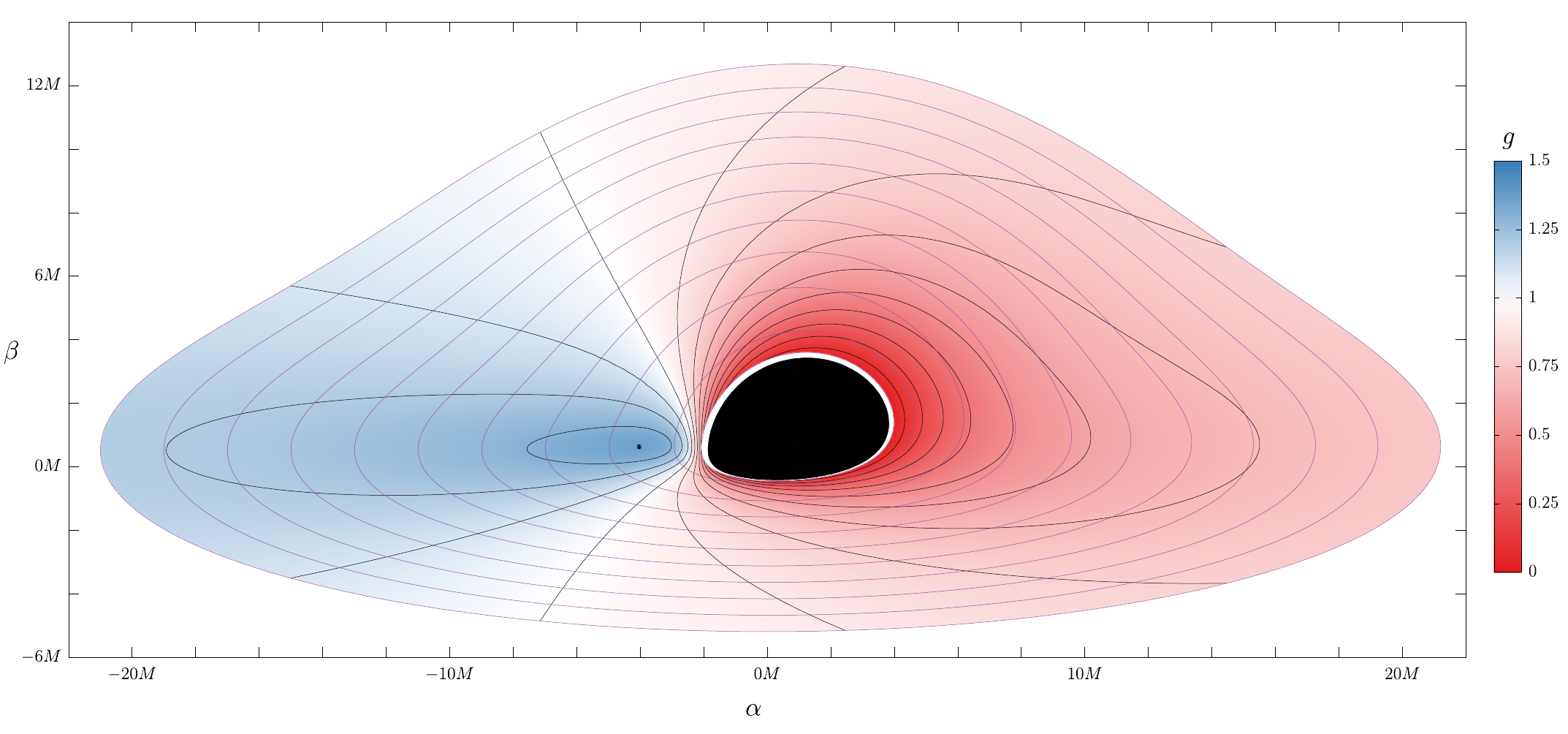}
	\caption{Direct ($\bar{m}=0$) image of an equatorial disk of isotropic emitters orbiting a Kerr black hole with spin $a/M=99.8\%$, viewed from an inclination angle $\theta_o=75^\circ$.  We display contours of fixed emission radius $r_s=2M,4M,\ldots,20M$ (purple) and contours of fixed observed redshift $g=0.1,0.2,\ldots1.3$ (black), providing a parameterization $(r_s,g)$ of the disk.  The dot indicates the source of the MOB in this image, $g\approx1.36$, which is less than the overall MOB (see Fig.~\ref{fig:LensedDisk}).  The inner edge of the disk is the ISCO $r_\mathrm{ms}$, and we colored the interior of the horizon contour $r_s=r_+$ black.  Similar plots appear in Ref.~\cite{Speith1993}.}
	\label{fig:DiskParameterization}
\end{figure*}

\begin{figure}
	\centering
	\includegraphics[width=\columnwidth]{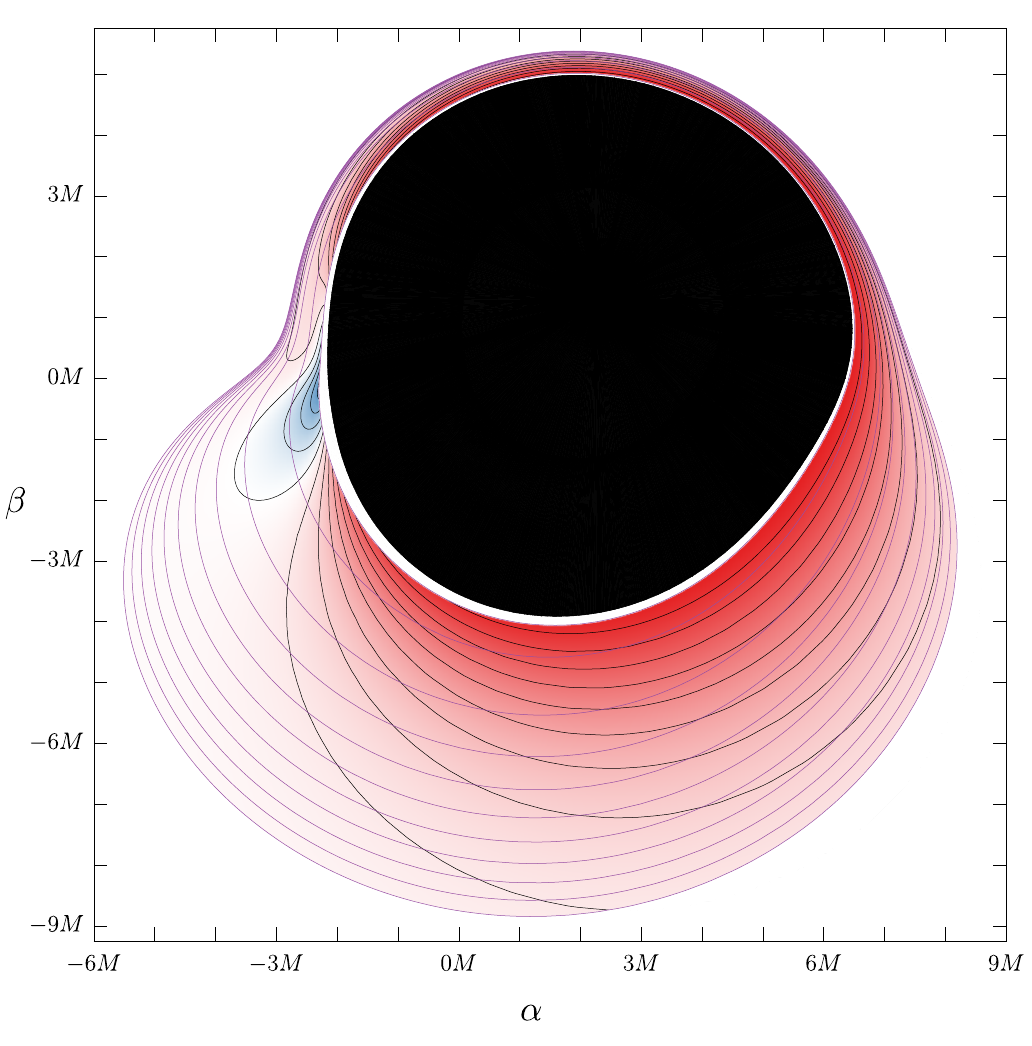}
	\caption{Lensed backside ($\bar{m}=1$) image of the equatorial disk in Fig.~\ref{fig:DiskParameterization}, with the same parameters and conventions.  The MOB from this image is the overall MOB, $g_\mathrm{mob}\approx1.43$.}
	\label{fig:LensedDisk}
\end{figure}

Bardeen \cite{Bardeen1973} introduced Cartesian coordinates on the image plane of an observer far from a Kerr black hole: a photon reaching a distant observer at inclination $\theta_o$ from the BH spin axis with conserved quantities $(\lambda,\eta)$ appears at Cartesian position $(\alpha,\beta)$, with
\begin{subequations}
\label{eq:BardeenCoordinates}
\begin{align}
	\alpha&=-\frac{\lambda}{\sin{\theta_o}},\\
	\beta&=\pm_o\sqrt{\eta+a^2\cos^2{\theta_o}-\lambda^2\cot^2{\theta_o}},
\end{align}
\end{subequations}
where $\pm_o=\sign\pa{p_o^\theta}$.  The optical appearance of an equatorial disk is determined by shooting light rays backwards into the geometry from every pixel $(\alpha,\beta)$ on the observer screen, keeping track of the point(s) of intersection with the disk.  A light ray emitted from $(\alpha,\beta)$ is traced by solving the null geodesic equation for the trajectory of the photon with corresponding parameters $(\lambda,\eta)$.  Since we will only consider a stationary, axisymmetric disk of prograde circular equatorial orbiters, we may ignore the toroidal components $(t,\phi)$ of the null geodesic equation.  Thus, we only need consider its poloidal $(r,\theta)$ component,
\begin{align}
	\label{eq:NullGeodesics}
	I_r=\fint_{r_s}^{r_o}\frac{\ed r}{\pm_r\sqrt{\mathcal{R}(r)}}
	=\fint_{\theta_s}^{\theta_o}\frac{\ed\theta}{\pm_\theta\sqrt{\Theta(\theta)}}
	=G_\theta,
\end{align}
where the slashes across the integral signs are meant to indicate that the signs $\pm_{r,\theta}$ switch at every radial and polar turning point, respectively, so that these geodesic path integrals grow secularly along the photon trajectory.

While a photon may only encounter at most one radial turning point along its trajectory, it may in principle execute arbitrarily many librations (polar oscillations) on its way from source to observer.  This is possible because the Kerr geometry admits spherical bound photon orbits, and photons with nearby conserved quantities may skirt these unstable orbits an arbitrarily long time.\footnote{This effect underlies the ``photon ring'' phenomenon \cite{Gralla2019,Johnson2019,Gralla2020a,Himwich2020,Gralla2020c}.}  As a result, an isotropically emitting equatorial source ring $r=r_s$ produces an infinite number of image rings on the screen of a distant observer, which differ by the number $m$ of polar turning points encountered by the corresponding photons between emission and observation.  As described in Sec.~VI of Ref.~\cite{Gralla2020a} (and illustrated in Fig.~6 therein), continuous curves (images of the source radius) on the observer screen are labeled not by $m$ but by an integer $\bar{m}=m-H(\beta)$, where $H$ denotes the Heaviside function.

Recent work on the null geodesic equation in the Kerr spacetime has led to its complete solution in terms of elliptic functions \cite{Gralla2020b}.  This allows us to carry out the ray-tracing necessary to describe the optical appearance of our disk model analytically.  In particular, the infinitely many images of an equatorial source ring of radius $r_s$ (as seen by a distant observer at inclination $\theta_o$) are obtained from Eq.~\eqref{eq:NullGeodesics} with $r_o\to\infty$ and $\theta_s=\pi/2$.  By specializing to equatorial sources, we can exclude vortical geodesics (with $\eta<0$) that can never reach the equator \cite{Kapec2020,Gralla2020b}.  In that case, the polar integral on the RHS evaluates to
\begin{align}
	\label{eq:AngularMinoTime}
	G_\theta^m&=\frac{1}{\sqrt{-u_-a^2}}\br{2mK\pa{\frac{u_+}{u_-}}-\sign\pa{\beta}F_o},\\
	F_o&=F\pa{\arcsin\pa{\frac{\cos{\theta_o}}{\sqrt{u_+}}}\left|\frac{u_+}{u_-}\right.},
\end{align}
where $u_\pm$ denote the roots of the angular potential, given in Eq.~(11) of Ref.~\cite{Gralla2020b} as
\begin{align}
	u_\pm=\triangle_\theta\pm\sqrt{\triangle_\theta^2+\frac{\eta}{a^2}},\quad
	\triangle_\theta=\frac{1}{2}\pa{1-\frac{\eta+\lambda^2}{a^2}},
\end{align}
while $m$ counts the number of polar turning points along the trajectory.  Using Eq.~\eqref{eq:BardeenCoordinates} to view $G_\theta^m$ as a function of $(\alpha,\beta,\theta_o)$, we find that it suffers from a discontinuity across the horizontal axis $\beta=0$.  Working with $\bar{m}$ instead of $m$ results in a smooth function $G_\theta^{\bar{m}}$.

For each $\bar{m}$, Eq.~\eqref{eq:NullGeodesics} admits a continuous set of solutions $(\lambda,\eta)$ that give rise, via Eq.~\eqref{eq:BardeenCoordinates}, to a continuous image of the source ring on the observer screen.  This $\bar{m}^\text{th}$ image of the disk is confined to the region of the image plane defined by the condition
\begin{align}
	\label{eq:LensingBand}
	G_\theta^{\bar{m}}<I_r^\mathrm{total}=
	\begin{cases} 
		\displaystyle2\int_{r_4}^{\infty}\frac{\ed r}{\sqrt{\mathcal{R}(r)}}&r_+<r_4\in\mathbb{R},\vspace{2pt}\\
		\displaystyle\int_{r_+}^{\infty}\frac{\ed r}{\sqrt{\mathcal{R}(r)}}&\text{otherwise}.
	   \end{cases}
\end{align}
These regions were called ``lensing bands'' in Ref.~\cite{Gralla2020c} and illustrated in Fig.~3 therein.  Closed-form expressions for the elliptic integral $I_r^\mathrm{total}$ are given in App.~A of Ref.~\cite{Gralla2020a}.

Inverting Eq.~\eqref{eq:NullGeodesics} for the source radius $r_s$ results in
\begin{align}
	\label{eq:EquatorialRadius}
	r_s^{(\bar{m})}(\alpha,\beta)=r_s(G_\theta^{\bar{m}}),
\end{align}
where (see Eq.~(30) of Ref.~\cite{Gralla2020a})
\begin{align}
	r_s(I_r)&=\frac{r_4r_{31}-r_3r_{41}\sn^2\pa{\frac{1}{2}\sqrt{r_{31}r_{42}}I_r-\mathcal{F}_o\big|k}}{r_{31}-r_{41}\sn^2\pa{\frac{1}{2}\sqrt{r_{31}r_{42}}I_r-\mathcal{F}_o\big|k}},\\
	\label{eq:k}
	\mathcal{F}_o&=F\pa{\left.\arcsin{\sqrt{\frac{r_{31}}{r_{41}}}}\right|k},\quad
	k=\frac{r_{32}r_{41}}{r_{31}r_{42}}.
\end{align}
Here, we introduced the notation
\begin{align}
	r_{ij}=r_i-r_j,\quad
	r_\pm=M\pm\sqrt{M^2-a^2},
\end{align}
with the roots $\cu{r_1,r_2,r_3,r_4}$ of the radial potential $\mathcal{R}(r)$ given in Ref.~\cite{Gralla2020b}.  The function $r_s^{(\bar{m})}(\alpha,\beta)$ is the $\bar{m}^\text{th}$ transfer function of the disk, with domain the $\bar{m}^\text{th}$ lensing band \eqref{eq:LensingBand}.  Additional details are provided in App~\ref{app:TransferFunctions}.

\subsection{Parameterizing an equatorial ring by redshift}

Each equatorial source ring $r=r_s$ in Kerr produces infinitely many images on the screen of a distant observer at inclination $\theta_o$ (see, e.g., Fig.~6 of Ref.~\cite{Gralla2020a}).  Each of these images is a closed curve labeled by an integer $\bar{m}$.  Let $\alpha_\mathrm{max/min}(r_s,\bar{m})$ respectively denote the abscissa of the rightmost/leftmost point produced on the observer screen by the $\bar{m}^\text{th}$ image of the source ring $r=r_s$.  By Eq.~\eqref{eq:BardeenCoordinates}, the maximal/minimal angular momentum of a photon in the $\bar{m}^\text{th}$ image of the source ring $r=r_s$ is
\begin{align}
	\label{eq:MaximalAngularMomentum}
	\lambda_\mathrm{max/min}(r_s,\bar{m})=-\alpha_\mathrm{min/max}\sin{\theta_o}.
\end{align}
Conversely, every value of $\lambda$ defines via  Eq.~\eqref{eq:BardeenCoordinates} a vertical line on the observer screen that intersects the $\bar{m}^\text{th}$ image of the source ring $r=r_s$ either: twice if $\lambda_\mathrm{min}<\lambda<\lambda_\mathrm{max}$, exactly once if $\lambda=\lambda_\mathrm{max/min}$, or else not at all.

We also define the maximal blueshift/redshift\footnote{In this equation, we specialized to emitters on prograde orbits.  For retrograde orbits, one must switch the definitions of $g_\mathrm{max/min}$.}
\begin{align}
	\label{eq:MaximumRedshift}
	g_\mathrm{max/min}(r_s,\bar{m})=g(r_s,\lambda_\mathrm{max/min}(r_s,\bar{m})).
\end{align}
By Eq.~\eqref{eq:RedshiftMonotonicity}, $g(r_s,\lambda)$ is monotonic in $\lambda$ at fixed $r_s$.  Hence, as $\lambda$ ranges over $\br{\lambda_\mathrm{min},\lambda_\mathrm{max}}$, the redshift $g$ achieves every value in $\br{g_\mathrm{min},g_\mathrm{max}}$ exactly once (and vice versa).

Thus, for a given equatorial ring $r=r_s$, each redshift $g\in\br{g_\mathrm{min}(\bar{m}),g_\mathrm{max}(\bar{m})}$ corresponds to exactly two points on the $\bar{m}^\text{th}$ image of the source ring, and accordingly to two source angles $\phi_s$ on the ring.  Hence, the angle around any equatorial source ring is parameterized by redshift, with the full ring obtained by sweeping over the range $\br{g_\mathrm{min},g_\mathrm{max}}$ twice.  This fact (that a source ring is a double cover of redshifts) was first noted by Cunningham \cite{Cunningham1975}, though he only considered the direct emission with $\bar{m}=0$.  We illustrate the parameterization $(r_s,g)$ of an equatorial disk by emission radius and observed redshift in Figs.~\ref{fig:DiskParameterization} and \ref{fig:LensedDisk}.

\section{Presence of the blue blade}
\label{sec:BlueBlade}

In this section, we show that for all inclinations $\theta_o$, the broadened spectrum of electromagnetic line emission exhibits the prominent ``blue blade'' displayed in Fig.~\ref{fig:BroadenedLine}.  More precisely, we argue that the observed flux retains a finite nonzero value up to the MOB, beyond which it must then (by definition) suddenly vanish.  The resulting discontinuity gives rise to a sharp feature: the blue blade.

\subsection{Cunningham's transfer function}

Cunningham's 1975 work \cite{Cunningham1975} introduced a transfer function to characterize the direct emission received by a far observer from a thin equatorial accretion disk of Kerr orbiters.  His approach has been reviewed several times since (see, e.g., Refs.~\cite{Dauser2010,Lupsasca2018}), so we only sketch it briefly here before using his technology to develop our argument farther below.  First, he noted that the specific flux $F_{E_o}(\theta_o)$ at energy $E_o$ carried to an observer at inclination $\theta_o$ by a bundle of photons is the product of its observed solid angle $\Omega$ and specific intensity $I_{E_o}$, i.e.,\footnote{The specific flux is the power (energy per time) transferred per unit area and per unit frequency ($\mathrm{W}\cdot\mathrm{m}^{-2}\cdot\mathrm{eV}^{-1}$).  The specific intensity (or spectral radiance) is the power transferred per unit area, per unit frequency, and per unit steradian ($\mathrm{W}\cdot\mathrm{m}^{-2}\cdot\mathrm{eV}^{-1}\cdot\mathrm{sr}^{-1}$).}
\begin{align}
	\ed F_{E_o}(\theta_o)=I_{E_o}\ed\Omega.
\end{align}
By Liouville's theorem on the invariance of the phase space density of photons, the specific intensity varies as the third power of the frequency along a beam of radiation \cite{Lindquist1966}.  It follows that the observed specific intensity $I_{E_o}$ is related to the emitted specific intensity $I_{E_s}$ by
\begin{align}
	I_{E_o}(\alpha,\beta)=g^3I_{E_s}(r_s,E_s=E_o/g).
\end{align}
A bundle of rays subtends a solid angle on the image
\begin{align}
	\label{eq:SolidAngle}
	\ed\Omega=\frac{1}{r_o^2}\ed\alpha\ed\beta
	=\frac{1}{r_o^2}\ab{\frac{\pd\pa{\alpha,\beta}}{\pd\pa{r_s,g^*}}}\ed r_s\ed g^*,
\end{align}
where we introduced the normalized redshift
\begin{align}
	g^*=\frac{g-g_\mathrm{min}}{g_\mathrm{max}-g_\mathrm{min}}
	\in\br{0,1}.
\end{align}
This enables us to express the specific flux at energy $E_o$ received by the observer as an area integral over the image of the disk, with its surface parameterized by $(r_s,g^*)$ as explained in the previous section.  This double integral over redshift and emission radius takes the form
\begin{align}
	\label{eq:SpecificFlux}
	F_{E_o}(\theta_o)&=\frac{1}{r_o^2}\int g^3I_{E_s}(r_s,E_o/g)\ed\alpha\ed\beta\\
	&=\frac{1}{r_o^2}\int\frac{\pi r_sg^2f(r_s,g^*)}{\sqrt{g^*\pa{1-g^*}}}I_{E_s}(r_s,E_o/g)\ed r_s\ed g^*,\nonumber
\end{align}
where we introduced Cunningham's transfer function
\begin{align}
	\label{eq:TransferFunction}
	f(r_s,g^*)=\frac{g\sqrt{g^*\pa{1-g^*}}}{\pi r_s}\ab{\frac{\pd\pa{\alpha,\beta}}{\pd\pa{r_s,g^*}}},
\end{align}
which is observer-dependent and whose particular form was chosen so that its numerical value varies less with $r_s$ and $g^*$ \cite{Cunningham1975}.  Recalling from the previous section that the parameterization of the disk by $(r_s,g^*)$ is double-valued, it follows that the redshift integral in Eq.~\eqref{eq:SpecificFlux} must be performed twice for each emission radius, once for each half $g^*\in\br{0,1}$ of the source ring.

Our Eq.~\eqref{eq:TransferFunction} exactly reproduces Eq.~(23) of Ref.~\cite{Dauser2010} and agrees with Cunningham's definition of the transfer function [Eq.~(5) of Ref.~\cite{Cunningham1975} with $r_o^2\pd\Omega$ replaced by $\pd\pa{\alpha,\beta}$ using Eq.~\eqref{eq:SolidAngle}].  His formula (7) for the ``effective luminosity'' $L=4\pi r_o^2F_{E_o}$ also agrees with our Eq.~\eqref{eq:SpecificFlux}.  However, note that Cunningham's transfer function $f(r_s,g^*)$ differs from the ``transfer functions'' $r_s^{(\bar{m})}(\alpha,\beta)$ defined in Eq.~\eqref{eq:EquatorialRadius} above, where we followed the nomenclature of Refs.~\cite{Gralla2019,Gralla2020a} instead.

\subsection{The transfer function is nonzero at the MOB}

\begin{figure}
	\centering
	\includegraphics[width=\columnwidth]{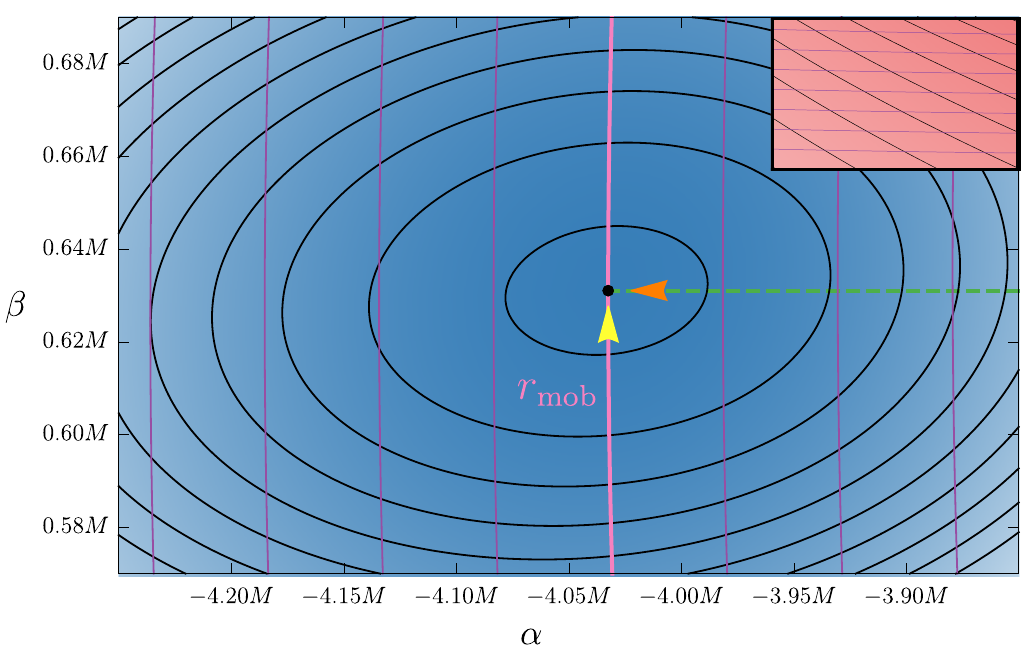}\qquad
	\caption{Zoomed-in version of Fig.~\ref{fig:DiskParameterization} showing the disk parameterization $(r_s,g^*)$ by emission radius and observed redshift near the MOB source $(r_\mathrm{mob},1)$, where the coordinate system becomes degenerate.  At generic points in the disk, the coordinate grid is regular (inset panel), but near the MOB, the redshift acts more like a radial coordinate with the MOB as origin.  Cunningham's transfer function $f(r_s,g^*)$ is multi-valued at $(r_\mathrm{mob},1)$, just like the function $\cos{\varphi}$ near the origin of the polar plane $(\rho,\varphi)$.  In the text, we consider two different limiting values as $g^*\to1$: one is obtained by taking the limit along the contour $r_s=r_\mathrm{mob}$ (yellow arrow), while the other is taken along the green horizontal line (orange arrow).}
	\label{fig:TransferFunction}
\end{figure}

In this section, we prove that Cunningham's transfer function \eqref{eq:TransferFunction} remains nonzero near the MOB, i.e.,
\begin{align}
	\label{eq:TransferFunctionMOB}
	\lim_{g^*\to1^-}f(r_\mathrm{mob},g^*)\neq0,
\end{align}
where $r_\mathrm{mob}$ denotes the source radius $r_s$ corresponding to emission with the maximum observed blueshift.  This emission appears on the observer screen at some position $(\alpha_0,\beta_0)$, where $\alpha_0=\alpha_\mathrm{min}(r_\mathrm{mob},\bar{m})$ for whichever value of $\bar{m}$ corresponds to the emission with largest blueshift.\footnote{Cunningham considered only direct emission with $\bar{m}=0$.  We will also neglect this parameter for the present argument, as it holds for all values of $\bar{m}$ (including that for which the MOB is achieved).}

Since $g^*(\alpha,\beta)$ achieves its maximum $g^*=1$ at $(\alpha_0,\beta_0)$, its expansion about that point must by definition be
\begin{align}
	\label{eq:RedshiftMOB}
	g^*(\alpha,\beta)\approx1&-\frac{g_1^2}{2}\pa{\alpha-\alpha_0}^2-\frac{g_2^2}{2}\pa{\beta-\beta_0}^2\nonumber\\
	&+g_3\pa{\alpha-\alpha_0}\pa{\beta-\beta_0}+\ldots,
\end{align}
for some constants $g_1,g_2,g_3$ such that $h=g_1^2g_2^2-g_3^2>0$, ensuring the Hessian matrix has positive determinant (as required for a local maximum).  Likewise, we must have
\begin{align}
	\label{eq:SourceRadiusMOB}
	r_s(\alpha,\beta)\approx r_\mathrm{mob}+\frac{b}{\sqrt{2}g_2}\pa{\alpha-\alpha_0}+\ldots,
\end{align}
for some constant $b$, with no term $\O{\beta-\beta_0}$ present at leading order since contours of fixed $r_s$ must be vertical at their leftmost point $\alpha_\mathrm{min}$ (see Fig.~6 of Ref.~\cite{Gralla2020a}).  These relations imply that at $(r_s,g^*)=(r_\mathrm{mob},1)$,
\begin{align}
	\label{eq:Jacobian}
	\ab{\frac{\pd\pa{\alpha,\beta}}{\pd\pa{r_s,g^*}}}_\mathrm{mob}=\frac{1}{\sqrt{b^2\pa{1-g^*}-h\pa{r_\mathrm{mob}-r_s}^2}}.
\end{align}
Plugging this Jacobian into Eq.~\eqref{eq:TransferFunction} then implies that
\begin{align}
	\label{eq:ExactTransferFunctionMOB}
	\lim_{g^*\to1^-}f(r_\mathrm{mob},g^*)=\frac{g_\mathrm{mob}}{\pi r_\mathrm{mob}\ab{b}}
	\neq0,
\end{align}
thereby proving Eq.~\eqref{eq:TransferFunctionMOB}, as desired.  A related result was claimed without explicit proof in Eq.~(28) of Ref.~\cite{Dauser2010}.  Note that the exact value of $f(r_s,g^*)$ at $(r_\mathrm{mob},1)$ depends on the direction that this point is approached from, with the formula \eqref{eq:ExactTransferFunctionMOB} valid only when approaching $g^*=1$ along the line $r_s=r_\mathrm{mob}$ (yellow arrow in Fig.~\ref{fig:TransferFunction}).

\subsection{The specific flux density is nonzero at the MOB}

We can rewrite the specific flux \eqref{eq:SpecificFlux} as an integral
\begin{align}
	F_{E_o}(\theta_o)=\int_0^1F_{E_o}(g^*,\theta_o)\ed g^*,
\end{align}
over the ``specific flux density''
\begin{align}
	\label{eq:FluxDensity}
	F_{E_o}(g,\theta_o)=\int\frac{g^2f(r_s,g^*)I_{E_s}(r_s,E_o/g)}{2r_o^2\sqrt{g^*\pa{1-g^*}}}\ed\pa{\pi r_s^2},
\end{align}
which is itself a radial integral along the relevant contour of fixed $g$, as parameterized by $r_s$ (Fig.~\ref{fig:DiskParameterization}).  Plotting this specific flux density produces broadened line profiles like those shown in Fig.~\ref{fig:BroadenedLine} (or Fig.~2 of Ref.~\cite{Dauser2010}), which display a blue blade.  We now argue that this feature arises from a discontinuity in the flux density,
\begin{align}
	\label{eq:FluxDiscontinuity}
	\lim_{g^*\to1^+}F_{E_o}(g,\theta_o)=0
	\neq\lim_{g^*\to1^-}F_{E_o}(g,\theta_o).
\end{align}

Although each of the quantities appearing in the integrand of Eq.~\eqref{eq:FluxDensity} are complicated functions of $(r_s,g,\theta_o)$, the argument itself is quite simple.  It relies on the nontrivial fact \eqref{eq:TransferFunctionMOB} together with the observation that only emission radii $r_s\approx r_\mathrm{mob}$ contribute to the MOB $g^*=1$.

More precisely, for a generic redshift $0\le g\le g_\mathrm{mob}$, the radial integral \eqref{eq:FluxDensity} is a line integral along the corresponding contour of fixed $g$; for the extremal value $g=g_\mathrm{mob}$, on the other hand, the contour degenerates to the point $(r_\mathrm{mob},1)$ on the disk (Fig.~\ref{fig:TransferFunction}).  We make sense of the resulting integral as the limiting value, obtained as $g^*\to1$, of the specific flux density evaluated at a nearby redshift $g\approx g_\mathrm{mob}$.  That is, we consider an infinitesimal neighborhood of $(r_\mathrm{mob},1)$ containing a contour of fixed $g^*=1-\delta^2$ for $0<\delta\ll1$.  This contour is parameterized by a small band of radii $r_s\approx r_\mathrm{mob}$ that give the corresponding range of integration in Eq.~\eqref{eq:FluxDensity}.  To determine this range, note from Eq.~\eqref{eq:SourceRadiusMOB} that within this small neighborhood of the MOB, $r_s(\alpha,\beta)$ varies fastest in the $\alpha$-direction (the direction of its gradient, corresponding to the horizontal green line in Fig.~\ref{fig:TransferFunction}).  Hence, by Eq.~\eqref{eq:RedshiftMOB} with $\beta=\beta_0$, a given $r_s\approx r_\mathrm{mob}$ can contribute to redshifts up to
\begin{align}
	\label{eq:CoordinateChange}
	g^*(r_s)\approx1-\pa{\frac{g_1g_2}{b}}^2\pa{r_s-r_\mathrm{mob}}^2+\ldots
\end{align}
Inversely, only emission radii
\begin{align}
	\ab{r_s-r_\mathrm{mob}}\lesssim c\delta,\quad
	c=\ab{\frac{b}{g_1g_2}},
\end{align}
contribute to the specific flux density at $g^*=1-\delta^2$, i.e., only these contours of fixed $r_s$ intersect the contour of $g^*=1-\delta^2$, which is therefore fully parameterized by radii in this band.  Hence, in this case, Eq.~\eqref{eq:FluxDensity} becomes
\begin{align}
	\label{eq:FluxLimit}
	F_{E_o}\pa{g(\delta),\theta_o}\stackrel{\delta\to0}{\approx}X\int_{r_\mathrm{mob}-c\delta}^{r_\mathrm{mob}+c\delta}
	\frac{\ed r_s}{\sqrt{b^2\delta^2-h\pa{r_\mathrm{mob}-r_s}^2}},
\end{align}
where we plugged in Eq.~\eqref{eq:TransferFunction} with its Jacobian \eqref{eq:Jacobian} evaluated at $g^*=1-\delta^2$, and introduced a nonzero constant
\begin{align}
	X=\frac{g_\mathrm{mob}^3}{r_o^2}I_{E_s}(r_\mathrm{mob},E_o/g_\mathrm{mob})
	\neq0,
\end{align}
as a proportionality factor.  After a change of variables to $u=(r_\mathrm{mob}-r_s)/(c\delta)$, we obtain the integral expression
\begin{align}
	F_{E_o}\pa{g(\delta),\theta_o}\stackrel{\delta\to0}{\approx}X\int_{-1}^1\frac{\ed u}{\sqrt{b^2/c^2-hu^2}},
\end{align}
which remains finite and nonzero as $\delta\to0$, evaluating to
\begin{align}
	\lim_{g^*\to1^-}F_{E_o}(g,\theta_o)=\frac{2X}{\sqrt{h}}\arctan\sqrt{\frac{h}{g_3^2}}
	\neq0.
\end{align}
Unlike in Eq.~\eqref{eq:ExactTransferFunctionMOB}, the $g^*\to1^-$ limit is taken along the orange arrow in Fig.~\ref{fig:TransferFunction}.  This proves Eq.~\eqref{eq:FluxDiscontinuity}, establishing the presence of the ``blue blade'' discontinuity.

\subsection{Comparison with Cunningham's work}

In Fig.~1b of Cunningham's 1975 paper \cite{Cunningham1975}, the curve labeled by $\cos{\theta_o}=0$ displays the redshift of photons received by an equatorial observer from an equatorial disk of emitters orbiting a black hole of spin $a/M=99.81\%$.  This curve peaks at a value slightly below $g=1.7$, whereas the true MOB in this case can be computed exactly using Eq.~\eqref{eq:ExactEquatorialMOB}, resulting in $g=1.718$.  The reason for this discrepancy is likely that Cunningham only considered the direct $\bar{m}=0$ emission, whereas for this spin the MOB is achieved by lensed $\bar{m}=1$ emission that executes a half-orbit around the black hole (see Figs.~\ref{fig:DiskParameterization} and \ref{fig:LensedDisk}).  For the almost identical spin $a/M=99.8\%$, whose MOB is plotted in green in Fig.~\ref{fig:MaximumBlueshift}, neglecting this effect would amount to using the dotted green curve ($\bar{m}=0$ emission from the top of the disk) instead of the solid green curve ($\bar{m}=1$ emission from the backside of the disk), which gives the overall MOB.

\section{Exact numerics}
\label{sec:Numerics}

\begin{figure}
	\centering
	\includegraphics[width=\columnwidth]{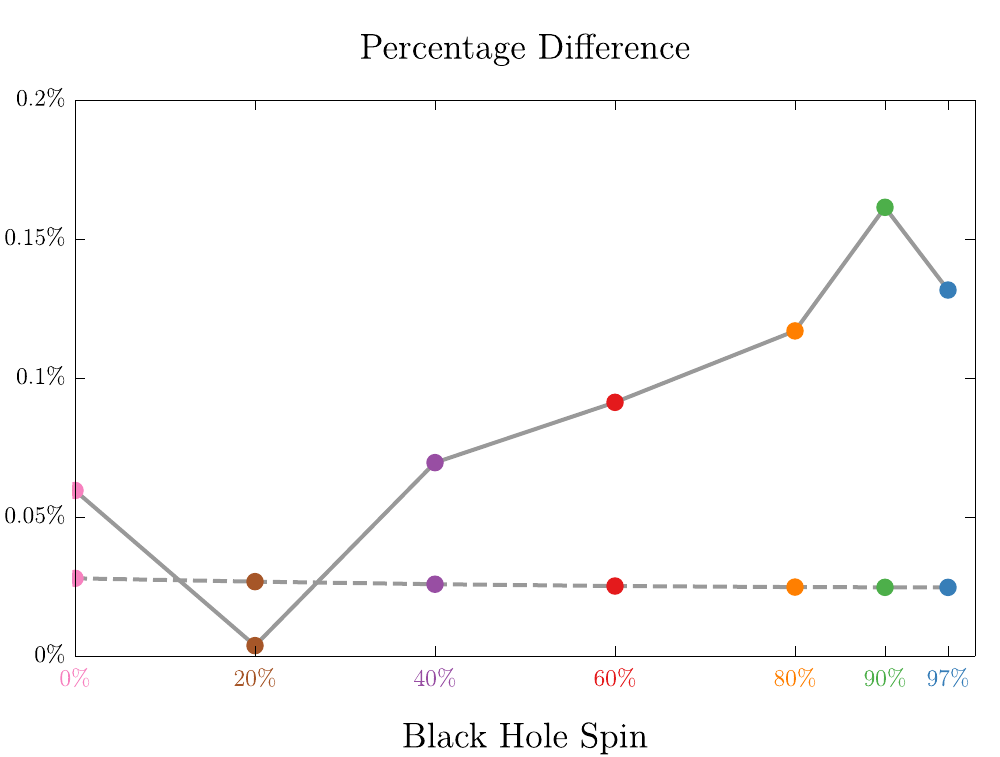}
	\caption{Percentage difference between the output of \texttt{relline} (Fig.~\ref{fig:BroadenedLine}) and our numerics (Fig.~\ref{fig:MaximumBlueshift}) for $g_\mathrm{mob}(a,\theta_o)$ [Eq.~\eqref{eq:MaximalBlueshift}] at inclinations $\theta_o=30^\circ$ (dashed) and $\theta_o=85^\circ$ (solid).}
	\label{fig:Comparison}
\end{figure}

\begin{figure*}
	\centering
	\includegraphics[width=\textwidth]{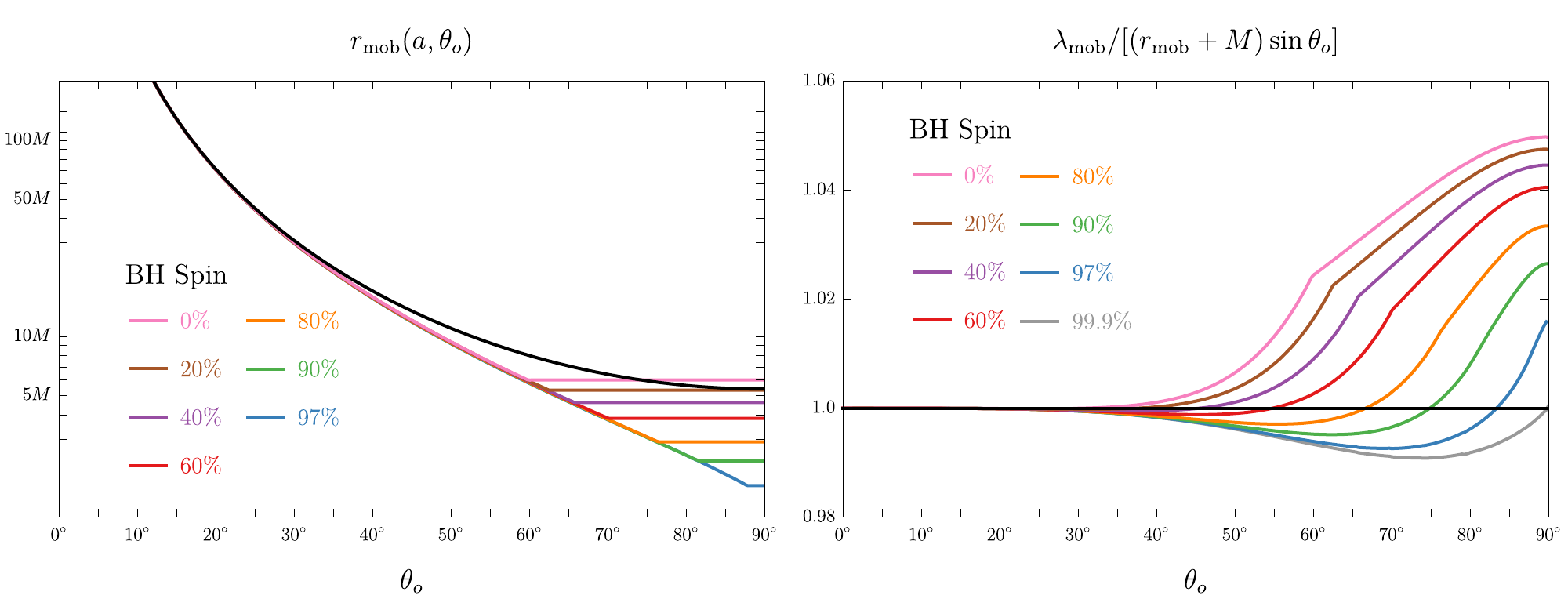}
	\caption{Left: Source radius $r_s=r_\mathrm{mob}(a,\theta_o)$ of the emission with maximum observed blueshift as a function of BH spin $a$ and observer inclination $\theta_o$.  Right: Ratio of the exact angular momentum $\lambda=\lambda_\mathrm{mob}(a,\theta_o)$ of the maximally blueshifted photons received by an observer at inclination $\theta_o$ as a function of BH spin $a$ to the analytic approximation $\lambda\approx\pa{r_s+M}\sin{\theta_o}$ [Eq.~\eqref{eq:ApproximateAngularMomentum}].  In both panels, the colored curves are exact and numerically computed following the procedure described in Sec.~\ref{sec:Numerics}, while the black curve corresponds to the analytic approximation developed for low inclinations in Sec.~\ref{subsec:LowInclinationAnySpin} below, i.e., using Eq.~\eqref{eq:AnalyticMaxRadius} for $r_\mathrm{mob}(a,\theta_o)$.  The right panel confirms the validity of these approximations at low inclinations.}
	\label{fig:Approximations}
\end{figure*}

In this section, we describe the numerical procedure that we followed to compute the exact value of the MOB for all inclinations and spins.  We then make a number of observations that will serve as the starting point of our analytic work in the next section.  As a check, we verify that the output of the code \texttt{relline} displayed in Fig.~\ref{fig:BroadenedLine} agrees with our own numerical results displayed in Fig.~\ref{fig:MaximumBlueshift}---see Fig.~\ref{fig:Comparison} for a quantitative comparison.

\subsection{The MOB as a maximum over disk images}

We have already defined $g_\mathrm{max}(r_s,\bar{m})$ to be the MOB received from light rays that encounter $\bar{m}$ angular turning points on their way to the observer after their emission from an equatorial source ring $r_s$.  Since our disk model consists of a collection of such source rings, the MOB from the entire disk is
\begin{align}
	\label{eq:MaximalBlueshift}
	g_\mathrm{mob}(a,\theta_o)=\max_{r_\mathrm{ms}\le r_s,0\le\bar{m}}g_\mathrm{max}(r_s,\bar{m}).
\end{align}
We analytically computed $g_\mathrm{mob}(a,\theta_o)$ for the extreme cases bracketing the range of possible inclinations: $\theta_o=0$ (polar observer) and $\theta_o=\pi/2$ (equatorial observer).

In the polar case, we found that $g_\mathrm{mob}(a,0)=1$, i.e., there can be no net blueshift.  This makes intuitive sense because the orbital motion of the matter in the disk is perpendicular to the line of sight to the polar observer, precluding any Doppler shift.  On the other hand, there is some gravitational redshift for all radii $r_s<\infty$, and hence the MOB---in this case, zero redshift---is achieved only by distant emission at $r_s\to\infty$.  In the equatorial case, on the other hand, the MOB arises in emission from the smallest radius in the disk: the ISCO radius $r_\mathrm{ms}$.  We compute the MOB $g_\mathrm{mob}(a,\pi/2)$ in Eq.~\eqref{eq:EquatorialMOB} below.

It seems very difficult (if at all possible) to extend the analytic computation to observers at other intermediate inclinations.  Nonetheless, it was helpful in our numerical analysis to know the exact values of $g_\mathrm{mob}(a,\theta_o)$ at both ends of the range of inclinations.

\subsection{Optimization of the maximization problem}

The computation of the MOB via the formula \eqref{eq:MaximalBlueshift} involves a maximization over two variables: since each source ring in the disk produces an infinite number of images on the observer screen, one must first determine which of these images labeled by $\bar{m}\ge0$ provides the largest contribution from each source radius; second, one must then maximize over all emission radii in the disk.

Accordingly, we split the maximization procedure into two steps.  First, we maximize over the blueshifts received from the $\bar{m}^\text{th}$ image of a source ring to obtain
\begin{align}
	\label{eq:RadialMaximization}
	g_\mathrm{mob}^{\bar{m}}(a,\theta_o)=\max_{r_\mathrm{ms}\le r_s}g_\mathrm{max}(r_s,\bar{m}),
\end{align}
and then we maximize over successive images to obtain
\begin{align}
	\label{eq:UnoptimizedMaximization}
	g_\mathrm{mob}(a,\theta_o)=\max_{0\le\bar{m}}g_\mathrm{mob}^{\bar{m}}(a,\theta_o).
\end{align}
We empirically observe that transitions from a regime in which the MOB arises from the $\bar{m}^\text{th}$ image to a regime in which it arises from the $\pa{\bar{m}+1}^\text{th}$ image always occur for the equatorial observer first.  This can be seen in the bottom panels of Fig.~\ref{fig:MaximumBlueshift}, where the transition from $\bar{m}=0$ to $\bar{m}=1$ appears as a kink in the solid curves plotting the MOB $g_\mathrm{mob}(a,\theta_o)$ as a function of $\theta_o$ (with the direct emission $\bar{m}=0$ plotted with dotted curves after the transition, where it no longer provides the overall MOB).  As the BH spin increases, the kink first appears at the rightmost edge corresponding to the equatorial observer.  Thus, for fixed BH spin, the largest value of $\bar{m}$ that we need to consider for any observer inclination, call it $\bar{m}^*$, is that which gives the MOB for the equatorial observer,
\begin{align}
	\label{eq:mMax}
	g_\mathrm{mob}^{\bar{m}^*}(a,\pi/2)=g_\mathrm{mob}(a,\pi/2),
\end{align}
which we know analytically [Eq.~\eqref{eq:EquatorialMOB} below].  We will prove this fact and give a simple formula for the spins where each transition $\bar{m}\to\bar{m}+1$ occurs in Sec.~\ref{sec:Analytics} below.  For the time being, this formula provides an effective bound $\bar{m}^*$ on the values of $\bar{m}$ that we need to maximize over, replacing Eq.~\eqref{eq:UnoptimizedMaximization} with its optimized variant
\begin{align}
	\label{eq:OptimizedMaximization}
	g_\mathrm{mob}(a,\theta_o)=\max_{0\le\bar{m}\le\bar{m}^*}g_\mathrm{mob}^{\bar{m}}(a,\theta_o).
\end{align}

\subsection{Description of the numerical procedure}

For each BH spin $a/M$, observer inclination $\theta_o$, and disk image $\bar{m}$, we compute $g_\mathrm{mob}^{\bar{m}}(a,\theta_o)$ as follows:
\begin{enumerate}
	\item Eq.~\eqref{eq:Redshift} gives the redshift $g(r_s,\lambda)$ of an observed photon as a function of its emission radius $r_s$ and energy-rescaled angular momentum $\lambda$.  Fixing a value of $\lambda$ defines a vertical line $\alpha=-\lambda\sin{\theta_o}$ on the image plane.  For every $\lambda$, we vary $\beta$ to find the minimum value achieved by $r_s^{(\bar{m})}(\alpha,\beta)$ along the corresponding line of fixed $\alpha$.  We denote this value by $r_\mathrm{min}(\lambda,\bar{m})$, as it is the smallest radius $r_s$ in the disk from which a photon with angular momentum $\lambda$ can be received in the $\bar{m}^\text{th}$ image of the disk.
	\item We invert $r_s=r_\mathrm{min}(\lambda,\bar{m})$ for $\lambda(r_s,\bar{m})$.  Since the point $(\alpha,\beta)$ where $r_s^{(\bar{m})}(\alpha,\beta)=r_\mathrm{min}(\lambda,\bar{m})$ is necessarily the leftmost point on the $\bar{m}^\text{th}$ image of the source ring $r_s$ [i.e., $\alpha=\alpha_\mathrm{min}(r_s,\bar{m})$], the resulting function is precisely $\lambda_\mathrm{max}(r_s,\bar{m})$ [Eq.~\eqref{eq:MaximalAngularMomentum}].
	\item We plug $\lambda=\lambda_\mathrm{max}(r_s,\bar{m})$ into $g(r_s,\lambda)$. By Eq.~\eqref{eq:RedshiftMonotonicity}, this yields the function $g_\mathrm{max}(r_s,\bar{m})$ [Eq.~\eqref{eq:MaximumRedshift}].
	\item We vary $r_s$ to find $g_\mathrm{mob}^{\bar{m}}(a,\theta_o)=g_\mathrm{max}(r_\mathrm{mob},\bar{m})$ and the associated emission radius $r_\mathrm{mob}(a,\theta_o)$.
\end{enumerate}
In practice, for any choice of spin $a/M$, we first consider the equatorial observer and compute $g_\mathrm{mob}^{\bar{m}}(a,\pi/2)$, starting with $\bar{m}=0$ and ending with the value $\bar{m}^*$ such that Eq.~\eqref{eq:mMax} is satisfied.  This determines the overall maximum $\bar{m}^*$ for all other observer inclinations.  As such, we only need to perform the radial maximization in Eq.~\eqref{eq:RadialMaximization} to compute $g_\mathrm{mob}^{\bar{m}}(a,\theta_o)$ for $\bar{m}\in\br{0,\bar{m}^*}$.  Finally, we maximize over $\bar{m}$ in this range to obtain the overall MOB $g_\mathrm{mob}(a,\theta_o)$ [Eq.~\eqref{eq:OptimizedMaximization}].

We carry out this procedure and display our results in Fig.~\ref{fig:MaximumBlueshift}.  In practice, we compute $g_\mathrm{mob}(a,\theta_o)$ for all the BH spins shown therein, sampling the observer inclination every degree for $10^\circ\le\theta_o\le50^\circ$ and then every $0.25^\circ$ for $50^\circ\le\theta_o\le89^\circ$, with a final computation carried out at $89.9^\circ$.  The MOB at the extreme inclinations $\theta_o=0^\circ$ and $90^\circ$ is known analytically, and it is essentially flat between $0^\circ$ and $10^\circ$.  As a practical matter, for $\theta_o\ge10^\circ$, it suffices to consider $\ab{\lambda}\lesssim50M$ and $\ab{\beta}\lesssim1.5M$ in step 1 above.  We show the corresponding values of $r_\mathrm{mob}(a,\theta_o)$ determined in step 4 in the left panel of Fig.~\ref{fig:Approximations}.

From the numerical computation described above, we glean information about the relationship between the MOB $g_\mathrm{mob}(a,\theta_o)$, its emission radius $r_\mathrm{mob}(a,\theta_o)$, and the label $\bar{m}(a,\theta_o)$ of the disk image from which it arises.  We can divide our observations about the MOB into four regimes of BH spin and observer inclination, as described in Sec.~\ref{sec:Summary}.  In the next section, we present (semi-)analytic approximations characterizing the MOB in each of these different regimes.

\section{Approximate (semi-)analytics}
\label{sec:Analytics}

Our numerics indicate that for relatively low observer inclinations $\theta_o$ ranging from $0^\circ$ (polar observer) to $\sim50^\circ$, the MOB always comes from the direct $\bar{m}=0$ image of the disk and is essentially independent of spin because it is sourced from radii far from the BH (Fig.~\ref{fig:Approximations} left).  This makes physical sense: at low inclinations, the direct light received from the disk is emitted upwards toward the observer (rather than beamed forward) and hence acquires little blueshift.  On the other hand, emission from the vicinity of the BH always incurs some gravitational redshift, and so the net observed blueshift arises from a large emission radius.  In the limiting case of a polar observer, whose line of sight to the BH is perpendicular to the accretion flow in the disk, the motion of the matter does not lend any blueshift to the direct emission.  As a result, the MOB is simply $g=1$, and it can only be achieved by emission from orbiters that are infinitely far from the BH, where there is no gravitational redshift.  Harnessing these insights, we analytically derive an excellent approximation to the MOB for those angles (Sec.~\ref{subsec:LowInclinationAnySpin}).

At higher inclinations, spin-dependent effects become important.  We observe that for fixed spin, the MOB increases monotonically in the inclination, reaching its maximal value for an equatorial observer.  This too aligns with physical intuition, as increasing the inclination enables the observer to receive more relativistically beamed photons, with significant blueshift imparted by the motion in the disk.  For an equatorial observer, the MOB always comes from ISCO emission and is monotonic in spin.  However, it does not always arise from the direct image of the ISCO.  We are able to analytically derive the MOB for an equatorial observer, as well as the label $\bar{m}$ of the ISCO image it appears in (Sec.~\ref{subsec:EquatorialObserverAnySpin}).  When the observer moves away from the equatorial plane, the monotonicity of the MOB in spin is broken.
 
In practice, we find that for ``low to moderate'' spins $0\le a/M\lesssim97\%$, the MOB arises from the direct $\bar{m}=0$ image for all observer inclinations $\theta_o$.  The low-inclination approximations developed for the MOB in Sec.~\ref{subsec:LowInclinationAnySpin} break down at ``high'' inclinations $\theta_o\gtrsim50^\circ$.  Nonetheless, for spins in the range $0\%\le a/M\le90\%$, we are able to fit a semi-analytic formula to our data that agrees with the exact numerics to better than $1\%$ precision (Sec.~\ref{subsec:HighInclinationLowSpin}).

Finally, for the case of both high spin ($a/M\gtrsim98\%$) and high inclination, we find that the MOB arises from strongly lensed images with $\bar{m}>0$.  Starting at $a/M\approx98\%$, near-equatorial observers can receive greater blueshifts from $\bar{m}=1$ images of the rings closest to the BH.  Physically, this phenomenon occurs because of a relativistic beaming effect: light rays can be slingshot around the BH and pick up a Doppler boost.  This is only possible for near-equatorial observers, with equatorial observers enjoying the largest possible blueshift.  As the BH approaches maximal spin, the dominant image with largest blueshift arises from increasingly strongly lensed images, with the label $\bar{m}$ (and consequently, the number of photon orbits executed around the BH) diverging logarithmically in the deviation $\kappa=\sqrt{1-a^2/M^2}\ll1$ from extremality (Sec.~\ref{subsec:HighInclinationHighSpin}).

\subsection{Low inclination, any spin}
\label{subsec:LowInclinationAnySpin}

We now develop an approximation that will enable us to derive an analytic formula for the MOB $g_\mathrm{mob}(a,\theta_o)$, valid at low inclinations $0^\circ\le\theta_o\lesssim50^\circ$ and for any value of the spin.  In this regime, our numerics reveal that the MOB always appears in the direct image of the disk, so we may focus on $\bar{m}=0$ trajectories of photons that never again cross the equatorial plane after emission.  That is, we focus on the transfer function $r_s^{(0)}(\alpha,\beta)$.

In the case of a precisely polar observer, the image of the disk is axisymmetric and parameterized by an impact radius $b=\sqrt{\alpha^2+\beta^2}$.  It was empirically noted in Eq.~(1) of Ref.~\cite{Gralla2020a} that the direct image of an equatorial source ring $r=r_s$ is itself a ring of constant radius
\begin{align}
	\label{eq:JustAddOne}
	b\approx r_s+M.
\end{align}
In other words, when $\theta_o=0^\circ$, the transfer function $r_s^{(0)}(b)$ for the direct image of an equatorial disk is simply
\begin{align}
	r_s^{(0)}\approx b-M.
\end{align}
We provide a proof of this relation in App.~\ref{app:TransferFunctions}.  The derivation proceeds from the exact expression for $r_s^{(0)}(b)$ given in Eq.~\eqref{eq:EquatorialRadius}, which may be systematically expanded in powers of $M/b$.  This results in the asymptotic expansion
\begin{align}
	\label{eq:PolarTransferFunction}
	r_s\stackrel{b\to\infty}\sim b-M+\frac{M^2-a^2}{2b}+\frac{3M^3\pa{5\pi-16}}{4b^2}+\ldots,
\end{align}
which improves as $b\to\infty$, or equivalently, as $r_s\to\infty$.  In the low-inclination regime that we are now considering, the MOB always arises from large-radius emission with $r_s\gg M$, and therefore $0<M/b\ll1$.  Hence, the subleading approximation \eqref{eq:JustAddOne} (``just add one'') will suffice for our present purposes.  We wish to emphasize that this remarkably simple formula is not a coordinate-independent statement; therefore, it should be regarded as a special property of the Boyer-Lindquist radius.

By Eq.~\eqref{eq:JustAddOne}, the circle $b\approx r_s+M$ on the polar observer screen is the direct image of the source ring $r=r_s$.  As the observer inclination increases, we would expect this circle to contract in the vertical direction.  (This is exactly what happens in flat space, where the image of an aligned circle is also a circle, while that of an inclined circle is an ellipse.)  The contour plots of $r^{(0)}(\alpha,\beta)$ depicted in Fig.~6 of \cite{Gralla2019} confirm this expectation: while the source rings' direct images are squeezed in the vertical direction, they maintain the same width (at least for small enough inclinations).  Therefore, the leftmost image produced by a source ring of radius $r_s$ on the screen of an observer at inclination $\theta_o$ ought to remain unaffected and still be
\begin{align}
	\alpha_\mathrm{min}\approx r_s+M.
\end{align}
As such, the maximal angular momentum $\lambda_\mathrm{max}(r_s,0)$ [Eq.~\eqref{eq:MaximalAngularMomentum} with $\bar{m}=0$] that an observer at low inclination $\theta_o$ can directly receive from a source ring of radius $r_s$ is
\begin{align}
	\label{eq:ApproximateAngularMomentum}
	\lambda_\mathrm{max}\approx-\pa{r_s+M}\sin{\theta_o}.
\end{align}
In principle, a better approximation could be obtained by keeping track of the subsubleading terms in Eq.~\eqref{eq:PolarTransferFunction}, but we will obtain an excellent approximation for the MOB even by neglecting them.  Note that Eq.~\eqref{eq:PolarTransferFunction} gives  $r_s\approx b$ to leading order (flat space), with the subleading $\O{b^0}$ correction $r_s\approx b-M$ accounting for the presence of a black hole via its mass only.  The BH spin $a$ enters the expansion only at subsubleading order $\O{b^{-1}}$; since we truncate our expansion at $\O{b^0}$, we expect to obtain a spin-independent approximation, as required to match the numerical results in this regime.  Physically, this spin-independence makes intuitive sense, as the light rays that account for the MOB are emitted far from the BH at large radii $r_s\gg M$, and are only weakly deflected as a result: the leading-order straight-line geodesics of flat spacetime are mildly bent by the gravitational pull of the mass $M$ (``monopole''), and its rotation $J=Ma$ (``dipole'') only affects null geodesics that pass much closer to the BH.

The analytic approximation \eqref{eq:ApproximateAngularMomentum} replaces step 2 in the numerical procedure outlined in Sec.~\ref{sec:Numerics} above.  Step 3 is carried out by plugging Eq.~\eqref{eq:ApproximateAngularMomentum} into $g(r_s,\lambda)$ to obtain
\begin{align}
	\label{eq:AnalyticRedshift}
	g_\mathrm{max}\approx\frac{\sqrt{r_s^3-3Mr_s^2+2a\sqrt{M}r_s^{3/2}}}{r_s^{3/2}+\sqrt{M}\br{a+\pa{r_s+M}\sin{\theta_o}}}.
\end{align}
This is the MOB of photons emitted from source radius $r_s$ that an observer at low inclination can directly receive  [Eq.~\eqref{eq:MaximumRedshift} with $\bar{m}=0$ and the upper choice of sign, corresponding to prograde orbiters].

Next, we must follow step 4 to find the radius $r_s=r_\mathrm{mob}$ for which this quantity is maximized.  That is, we must solve $g_\mathrm{max}'(r_s)=0$ for $r_s=r_\mathrm{mob}$.  The exact expression for $g_\mathrm{max}'(r_s)$ is rather complicated, but since we are only interested in $r_s\gg M$, we may considerably simplify it by expanding it in a small parameter $0\ll M/r_s\ll1$.  For a polar observer, this yields
\begin{align}
	g_\mathrm{max}'(r_s)=\frac{3}{2r_s^2}+\O{r_s^{-5/2}},
\end{align}
which implies that $r_\mathrm{mob}=\infty$, with a corresponding MOB $g_\mathrm{mob}(a,0)=1$, as mentioned before.  For a non-polar observer, we instead obtain the condition
\begin{align}
	\label{eq:AnalyticMaxRadiusCondition}
	0=xr_s^3+3\sqrt{M}r_s^{5/2}+3Mxr_s^2+\O{r_s^{3/2}},
\end{align}
which is manifestly spin-independent.  As in Eq.~\eqref{eq:ApproximateAngularMomentum}, the spin-dependent ``dipole'' terms only come in at higher order in the expansion.

The root $r_s=r_\mathrm{mob}$ of Eq.~\eqref{eq:AnalyticMaxRadiusCondition} is
\begin{align}
	r_\mathrm{mob}=\frac{3M}{2x^2}\pa{3-2x^2+\sqrt{3}\sqrt{3-4x^2}},
\end{align}
which is only defined for $0\le x\le\sqrt{3}/2$, that is, for angles $\theta_o\le60^\circ$.  As we are working at low inclination, we may replace this expression by its expansion in small $x$,
\begin{align}
	\label{eq:MaxRadiusExpansion}
	r_\mathrm{mob}\approx\frac{9M}{x^2}-6M+\O{x},
\end{align}
which is justified because this expression receives $\O{x}$ corrections once the $\O{r_s^{3/2}}$ subleading terms are included in Eq.~\eqref{eq:AnalyticMaxRadiusCondition}.  Hence, any expression $r_\mathrm{mob}(x)$ with the same small-$x$ expansion is equally valid at this order in perturbation theory.  We will use
\begin{align}
	\label{eq:AnalyticMaxRadius}
	r_\mathrm{mob}\approx\frac{9M}{x^2\pa{1+\frac{6}{9}x^2}},
\end{align}
which includes additional higher-order terms in $x$ chosen in order to improve the approximation at higher inclinations.  This can be seen in Fig.~\ref{fig:Approximations}, where we plot this approximate formula for $r_\mathrm{mob}$ (left panel), together with the resulting approximation for $\lambda_\mathrm{max}(r_\mathrm{mob})$ obtained by plugging Eq.~\eqref{eq:AnalyticMaxRadius} into Eq.~\eqref{eq:ApproximateAngularMomentum} (right panel), against their numerically computed exact values.

Finally, the MOB $g_\mathrm{mob}(a,x)$ is obtained by plugging Eq.~\eqref{eq:AnalyticMaxRadius} into Eq.~\eqref{eq:AnalyticRedshift}.  While the resulting function formally depends on the BH spin, we expect this spin-dependence to be weak because of the spin-independence of our approximations in Eqs.~\eqref{eq:ApproximateAngularMomentum} and \eqref{eq:AnalyticMaxRadius}.  Indeed, the first three terms in the small-$x$ expansion of $g_\mathrm{mob}(a,x)$ are spin-independent, and so plotting it as a function of $a$ produces curves that are almost identical everywhere they match the numerical data, as expected.  Hence, to simplify our final expression, we may as well set $a=0$, resulting in the expression given in Eq.~\eqref{eq:LowInclination} above,
\begin{align}
	\label{eq:LowInclinationMOB}
	g_\mathrm{mob}(a,x)\approx G(x).
\end{align}
Fig.~\ref{fig:MaximumBlueshift} shows excellent agreement at low inclinations between this approximation (black curve) and the numerically computed exact MOB at all spins (colored curves).

\subsection{Equatorial observer, any spin}
\label{subsec:EquatorialObserverAnySpin}

In this section, we analytically compute the MOB that can be radiated to the celestial sphere from an equatorial disk terminating at the ISCO.  We find that the maximally blueshifted photons are always received by equatorial observers.

As in Refs.~\cite{Cunningham1973,Gralla2017}, we parameterize emission from the disk using angles $(\Theta,\Psi)$ defined as the direction cosines of photons in the local rest frame \eqref{eq:OrbiterFrame} of each orbiter,
\begin{align}
	\cos{\Theta}&=-\frac{p^{(\theta)}}{p^{(t)}}
	=\mp\frac{g\sqrt{\eta}}{r_s},\\
	\label{eq:ForwardAngle}
	\cos{\Psi}&=\frac{p^{(\phi)}}{p^{(t)}}
	=\frac{\omega r_s\sqrt{\Delta(r_s)}\lambda-2aMv_s\pa{1-\omega\lambda}}{2aM\pa{1-\Omega_s\lambda}},
\end{align}
where $\omega$ [defined in Eq.~\eqref{eq:FrameDragging} above] is to be evaluated at $(r,\theta)=(r_s,\pi/2)$, while $\Omega_s$ and $v_s$ were given in Eqs.~\eqref{eq:AngularVelocity}--\eqref{eq:LNRF}.  Here, $p^{(a)}=\mathbf{e}_\mu^{(a)}p^\mu$ are the components of the photon momentum \eqref{eq:PhotonMomentum} in the local frame of the emitter, and the sign of $\cos{\Theta}$ must be chosen according to that of $p^{(\theta)}$ at the source.  Inverting the last relation for $\lambda(\Psi)$ and plugging into $g(r_s,\lambda)$ [Eq.~\eqref{eq:Redshift} with the $+$ choice of sign for prograde orbiters] yields the observed redshift of photons received at infinity as a function of their emission radius $r_s$ and local emission angle $\Psi$,
\begin{align}
	g=\frac{r_s^{3/2}-2M\sqrt{r_s}+\sqrt{M}\pa{a+\sqrt{\Delta(r_s)}\cos{\Psi}}}{\sqrt{r_s^3-3Mr_s^2+2a\sqrt{M}r_s^{3/2}}}.
\end{align}
For any given emission radius, this redshift is maximized by forward emission in the $+\phi$-direction with $\Psi=0$,
\begin{align}
	g_\phi(r_s)=\frac{r_s^{3/2}-2M\sqrt{r_s}+\sqrt{M}\pa{a+\sqrt{\Delta(r_s)}}}{\sqrt{r_s^3-3Mr_s^2+2a\sqrt{M}r_s^{3/2}}}.
\end{align}
One can show that $g_\phi(r_s)$ is monotonically decreasing,
\begin{align}
	r_s\ge r_\mathrm{ms}\quad\Longrightarrow\quad
	\pd_{r_s}g_\phi(r_s)<0.
\end{align}
It thus follows that the ``bluest'' photons emitted from the disk originate from the ISCO $r_s=r_\mathrm{ms}$, which is the smallest possible emission radius in the disk.  Moreover, since these photons are emitted from and into the equatorial plane, they must stay there forever by equatorial reflection symmetry.\footnote{This can be proved by noting that $\Psi=0$ requires $\Theta=\pi/2$, which in turn implies $\eta=0$ and hence $p^\theta=0$ in Eq.~\eqref{eq:PhotonMomentum}.}  Hence, they reach $\theta_o=90^\circ$, where
\begin{align}
	\label{eq:EquatorialMOB}
	g_\mathrm{mob}(a,x=1)=g_\phi(r_\mathrm{ms})
	=H(a),
\end{align}
as claimed in Eq.~\eqref{eq:ExactEquatorialMOB} above.  These photons have angular momentum $\lambda$ given by Eq.~\eqref{eq:ForwardAngle} with $\Psi=0$ and $r_s=r_\mathrm{ms}$,
\begin{align}
	\label{eq:AngularMomentumMOB}
	\lambda&=a+\frac{r_\mathrm{ms}^{3/2}}{\sqrt{M}}-\frac{\sqrt{r_\mathrm{ms}^3-3Mr_\mathrm{ms}^2+2a\sqrt{M}r_\mathrm{ms}^{3/2}}}{\sqrt{M}H(a)}.
\end{align} 
We will denote this quantity $\lambda_\mathrm{mob}(a)$, suppressing the argument $x=1$.  As explained in Sec.~\ref{sec:Numerics}, it is also equal to the function $\lambda_\mathrm{max}(r_s,\bar{m})$ [defined in Eq.~\eqref{eq:MaximalAngularMomentum} above] evaluated at the source radius $r_s=r_\mathrm{ms}$ of the MOB,
\begin{align}
	\label{eq:AngularMomentumISCO}
	\lambda_\mathrm{mob}(a)=\lambda_\mathrm{max}(r_\mathrm{ms},\bar{m}^*),
\end{align}
which by definition appears in the disk image $\bar{m}=\bar{m}^*$.

This relation implicitly defines a function $\bar{m}^*(a)$ that determines the maximum number of angular turning points encountered by the most blueshifted photons emitted from the disk on their way from source to observer.  This function is useful to know as it determines the range of $\bar{m}$ to be maximized over in our numerical procedure [Eq.~\eqref{eq:OptimizedMaximization} above].  Unfortunately, while the LHS \eqref{eq:AngularMomentumMOB} is known analytically, the function $\lambda_\mathrm{max}$ on the RHS is not (though its argument $r_\mathrm{ms}$ is).

We may nonetheless proceed as follows.  First, note that for an equatorial observer ($\theta_o=90^\circ$), the BH+disk model is symmetric under reflections about the equator, as neither the disk nor the observer inclination break this underlying symmetry of the Kerr metric.  This reflection symmetry in the geometry in turn guarantees a reflection symmetry in images of the disk.  In particular, images produced on the observer screen by a source ring $r=r_s$ must be symmetric about the $\alpha$-axis, with their leftmost point $\alpha_\mathrm{min}(r_s,\bar{m})=-\lambda_\mathrm{max}(r_s,\bar{m})$ [Eq.~\eqref{eq:MaximalAngularMomentum} with $\theta_o=90^\circ$] always appearing on this axis, i.e., with $\beta=0$.  Hence, for an equatorial observer, the transfer functions \eqref{eq:EquatorialRadius} evaluated at the leftmost point of a ring image obey
\begin{align}
	r_s^{(\bar{m})}\pa{-\lambda_\mathrm{max}(r_s,\bar{m}),0}=r_s,
\end{align}
Together with Eqs.~\eqref{eq:AngularMomentumMOB}--\eqref{eq:AngularMomentumISCO}, this implies that
\begin{align}
	r_s^{(\bar{m}^*)}\pa{-\lambda_\mathrm{mob}(a),0}=r_\mathrm{ms}.
\end{align}
Every quantity appearing in this equation is now known analytically, providing an effective way to compute $\bar{m}^*(a)$ for any given BH spin as the smallest integer value of $\bar{m}\ge0$ such that this relation holds.

Using this method, we can easily determine the spins $a=M\sqrt{1-\kappa^2}$ for which the MOB transitions from the $\bar{m}^\text{th}$ to the $(\bar{m}+1)^\text{th}$ image.  These transitions occur at:
\begin{subequations}
\label{eq:Transitions}
\begin{align}
	\bar{m}(0\to1):\qquad&\kappa\approx0.2045,\\
	\bar{m}(1\to2):\qquad&\kappa\approx1.713\times10^{-3},\\
	\bar{m}(2\to3):\qquad&\kappa\approx1.536\times10^{-5},\\
	\bar{m}(3\to4):\qquad&\kappa\approx1.381\times 10^{-7},\quad\ldots
\end{align}
\end{subequations}
That is, the equatorial MOB arises from direct $\bar{m}=0$ emission off the top of the disk for spins $a/M\le97.88\%$, and from lensed $\bar{m}=1$ emission off the backside of the disk when $97.88\%\le a/M\le99.99985\%$.  Thereafter, $\bar{m}^*$ increases by 1 roughly every time $\kappa=\sqrt{1-(a/M)^2}$ is divided by 100, or equivalently, for every four nines added to the spin $a/M=0.999\ldots$, in agreement with the near-extremal behavior first noticed in the high-spin expansion of Ref.~\cite{Gralla2017}.  It was shown therein that, for photons connecting the ISCO of a near-extremal black hole to infinity, the number of angular turning points diverges logarithmically in the deviation from extremality, i.e., $\bar{m}^*\sim-\log{\kappa}$.  We therefore fit the model $\bar{m}_0(\kappa)=c_0-c_1\log{\kappa}+c_2\kappa$ to the data \eqref{eq:Transitions}, and we obtain an excellent fit with
\begin{align}
	\label{eq:mNumerics}
	\bar{m}_0(\kappa)\approx0.6482-0.2122\log{\kappa}+0.07347\kappa.
\end{align}
This model correctly reproduces the spins at which the transitions \eqref{eq:Transitions} occur via
\begin{align}
	\label{eq:DivergentTurningPoints}
	\bar{m}^*(\kappa)=\left\lfloor\bar{m}_0(\kappa)\right\rfloor.
\end{align}
Remarkably, the best-fit parameters in Eq.~\eqref{eq:mNumerics} are in excellent agreement with the analytic formula (4.3) in Ref.~\cite{Gralla2017}, which predicts for these equatorial light rays\footnote{From Ref.~\cite{Gralla2017}, we take (4.3) and set $\hat{G}_\theta=0$ for an equatorial observer (App.~D), $D_o\sim R_o\to\infty$ for a far observer [Eq.~(3.20)], $\bar{R}=2^{1/3}$ for ISCO emission [Eq.~(3.5)], $\epsilon=\kappa^{2/3}$ [Eq.~(3.2)], $q=\sqrt{3}$ for equatorial rays [Eq.~(3.13)], and hence $qG_\theta=\pi$ (App.~D).}
\begin{align}
	m_0&=\frac{1}{\pi}\log\pa{\frac{18\times2^{2/3}}{2+\sqrt{3}}}-\frac{2}{3\pi}\log{\kappa}\\
	&\approx0.6479-0.2122\log{\kappa}.
\end{align}
The perfect agreement with Eq.~\eqref{eq:mNumerics} simultaneously lends credence to both our numerics and the analytic results obtained in Ref.~\cite{Gralla2017}.

This concludes the analytic derivation of the exact spin-dependent MOB received by an equatorial observer [Eq.~\eqref{eq:ExactEquatorialMOB}], and of the analytic approximation to the maximal $\bar{m}^*$ of disk images relevant to the computation of the MOB [Eq.~\eqref{eq:MaxTurningPoints}].  In Fig.~\ref{fig:Approximations}, the equatorial values of the MOB for the considered spins are illustrated with dots along the axis $\theta_o=90^\circ$.  The numerically calculated MOB curves terminate on those dots, verifying that we used the correct value of $\bar{m}^*$ for each case.

\subsection{High inclination, low to moderate spin}
\label{subsec:HighInclinationLowSpin}

In the regime of high inclination and low-to-moderate spin, the source radius of the MOB moves in closer to the black hole.  As a result, spin effects become important and the MOB acquires a significant spin-dependence.  For these reasons, the analytic approximations developed in Sec.~\ref{subsec:LowInclinationAnySpin} above break down.  Nonetheless, by fitting simple models to our numerical data, we obtain semi-analytic approximations that reproduce the curves in Fig.~\ref{fig:BroadenedLine} with high fidelity over the entire range of spins $0\le a/M\le97\%$, in which only the direct $\bar{m}=0$ emission is relevant to the MOB.

To develop a good semi-analytic model for the MOB, we start with $G(x)$ [Eq.~\eqref{eq:LowInclination}], which gives an excellent approximation for low inclination and any spin.  We combine this with $H(a)$, which gives the analytic result for the maximal inclination $\theta_o=90^\circ$ and any spin, and examine the difference at intermediate spins as encoded in the quantity $\chi(a,x=\sin{\theta_o})$, defined in Eq.~\eqref{eq:SemiAnalyticFit} above by
\begin{align}
	g_\mathrm{mob}(a,x)=G(x)+\br{H(a)-G(1)}\chi(a,x).
\end{align}
Since $g_\mathrm{mob}(a,0)=G(0)$ and $g_\mathrm{mob}(a,1)=H(a)$, we need
\begin{align}
	\lim_{x\to0}\chi(a,x)=0,\quad
	\lim_{x\to1}\chi(a,x)=1.
\end{align}
In practice, we find that the semi-analytic fit
\begin{align}
	\label{eq:HighInclinationMOB}
	\chi=xe^{-\frac{\pa{3+7a}\pa{1-x^2}}{\sqrt{1-a^2}}},
\end{align}
agrees with the numerical data to within 1\% from Schwarzschild ($a=0$) to spin $a/M=90\%$.  It also fits the data up to $a/M=97\%$ to within 4\% accuracy.  Alternatively, the semi-analytic fit
\begin{align}
	\label{eq:AlternativeFit}
	\chi=x\pa{1-\erf\br{\pa{\frac{5}{4}+\frac{7}{10}\frac{a}{\sqrt{1-a^2}}}\sqrt{1-x^2}}}
\end{align}
matches the data to within 1.5\% accuracy over the entire range $0\le a/M\le 97\%$.  Finally,
\begin{align}
	\chi=x\pa{1-\erf\br{\frac{3}{2}\pa{1+\frac{a}{\sqrt{1-a^2}}}\pa{1-x^2}^{2/3}}}
\end{align}
fits the data to within 1\% for spins $0\le a/M\le 90\%$ and to within 2.5\% for spins $90\%\le a/M\le 97\%$.

\subsection{High inclination, high spin}
\label{subsec:HighInclinationHighSpin}

For observers at high inclinations, the MOB arises from ISCO emission.  As the black hole spin approaches extremality ($a\to M$), the ISCO radius approaches the event horizon radius, $r_\mathrm{ms}\to r_+=M$.  At first sight, this is problematic because the ISCO is a timelike geodesic, whereas the horizon is a null hypersurface.

This apparent paradox was first resolved by Bardeen, Press and Teukolsky \cite{Bardeen1972}, who observed that while the Boyer-Lindquist coordinate distance $r_\mathrm{ms}-r_+$ shrinks to zero as $a\to M$, the radial proper distance between these radii diverges logarithmically in the deviation of the BH from extremality $0\le\kappa=\sqrt{1-a^2/M^2}\ll1$.  This gives rise to a throat-like region outside the event horizon of the black hole (illustrated in Fig.~2 of Ref.~\cite{Bardeen1972}), which was much later understood by Bardeen and Horowitz \cite{Bardeen1999} to form a nondegenerate vacuum solution to the Einstein equations in its own right.  This spacetime, which was dubbed the Near-Horizon Extreme Kerr (NHEK) geometry, exhibits many interesting features (such as conformal symmetry) and connections to quantum gravity (e.g., via the Kerr/CFT correspondence \cite{Guica2009})---see Sec.~III of Ref.~\cite{Kapec2020} for a recent astrophysically oriented review.

Since the ISCO lies in the throat region that develops near extremality, it follows that at high spin, the MOB received by near-equatorial observers arises from emission in the NHEK geometry.  Light rays connecting the throat region to infinity simplify significantly thanks to the enhanced symmetry of the NHEK geometry \cite{Porfyriadis2017,Gralla2017}, which enjoys an enlarged $\mathsf{SL}(2,\mathbb{R})\times\mathsf{U}(1)$ isometry group.

In particular, consider a BH with near-extremal spin $a=M\sqrt{1-\epsilon^3}$, where $0<\epsilon\ll1$ is a small deviation from extremality.  Photons emitted from orbiters in the NHEK region that emerges as $\epsilon\to0$ must have angular momentum of the form (Eq.~(3.7) of Ref.~\cite{Gralla2017})
\begin{align}
	\label{eq:AngularMomentumExpansion}
	\lambda=2M\pa{1-\epsilon\bar{\lambda}},
\end{align}
for some $\bar{\lambda}\sim\O{\epsilon^0}$.  This condition is necessary to ensure they have finite energy in the local rest frame \eqref{eq:OrbiterFrame} of the emitter.  This implies that images of a source ring
\begin{align}
	r_s=M\pa{1+\epsilon\bar{R}_s}
\end{align}
of constant radius $\bar{R}$ in the throat must appear on the screen of a distant observer near the so-called NHEKline,
\begin{align}
	\label{eq:NHEKline}
	\alpha=-\frac{2M}{\sin{\theta_o}},\quad
	\ab{\beta}<M\sqrt{3+\cos^2{\theta_o}-4\cot^2{\theta_o}},
\end{align}
derived in App.~A of Ref.~\cite{Gralla2017}.  The NHEKline's height shrinks as the observer moves out of the equatorial plane, eventually vanishing altogether beyond the critical angle $\theta_c=\arctan\pa{4/3}^{1/4}\approx47^\circ$.  In the throat, the ISCO lies at NHEK radius (Eq.~(3.3) of Ref.~\cite{Gralla2017})
\begin{align}
	\bar{R}_s=2^{1/3}.
\end{align}

In this regime, for each number $m$ of angular turning points, the geodesic equation \eqref{eq:NullGeodesics} can be analytically solved for $\bar{\lambda}(\eta,m)$, resulting in (Eq.~(3.36) of Ref.~\cite{Gralla2017})
\begin{align}
	\label{eq:GeodesicsNHEK}
	\bar{\lambda}=\frac{2\Upsilon_m(q)}{4-q^2}\br{2-q\sqrt{1+\frac{\bar{R}}{2\Upsilon_m(q)}}\pa{4-q^2}},
\end{align}
where (inverting the definition $\eta=M^2(3-q^2)\ge0$)
\begin{align}
	\label{eq:Upsilon}
	\Upsilon_m(q)=\frac{q^4}{q+2}\frac{e^{-qG_\theta^mM}}{\epsilon},\quad
	q=\sqrt{3-\frac{\eta}{M^2}}\ge0,
\end{align}
with the geodesic integral $G_\theta^m$ as given in Eq.~\eqref{eq:AngularVelocity} and evaluated at $a=M$ and $\lambda=2M$.

Plugging $\bar{\lambda}(\eta,m)$ into the $\phi$-component of the geodesic equation gives the azimuth swept by a photon in terms of its conserved quantity $q$ as (Eq.~(3.41) of Ref.~\cite{Gralla2017})
\begin{align}
	\label{eq:SourceAngle}
	\Delta\phi\approx\frac{1}{2\bar{\lambda}\epsilon}\pa{\frac{\sqrt{q^2\bar{R}^2+8\bar{\lambda}\bar{R}+4\bar{\lambda}^2}}{\bar{R}}-q}.
\end{align}
Hence, $q$ (or $\eta$) parameterizes the angle $\phi_s\equiv-\Delta\phi(q)\mod2\pi$ around the source ring $R_s=\bar{R}_s$.  However, since $\Delta\phi(q)$ diverges linearly as $\epsilon\to0$, its range exceeds $2\pi$ in this limit.  Therefore, this parameterization is rather peculiar, as it covers the source ring multiple times!

That is, each source ring $\bar{R}=\bar{R}_s$ produces multiple image rings labeled by $\bar{m}$, and for each such image ring, the portion near the NHEKline ``unwinds'' the source ring multiple times on the observer screen.  Physically, this occurs because the light rays connecting the source ring to the observer execute increasingly many windings in $\phi$ per libration in $\theta$ as $\epsilon\to0$.

Likewise, since $a=M$ and $\lambda=2M$ to leading order in $\epsilon$, Eqs.~\eqref{eq:BardeenCoordinates} show that $\beta^2=\eta+M^2\cos^2{\theta_o}-4M^2\cot^2{\theta_o}$.  As such, using Eq.~\eqref{eq:Upsilon} to interchange $q$ and $\eta$, we have
\begin{align}
	\label{eq:NHEKlinePosition}
	q=\sqrt{3-\frac{\beta^2}{M^2}+\cos^2{\theta_o}-4\cot^2{\theta_o}}.
\end{align}
Thus, varying the vertical position $\beta$ along the NHEKline also varies the angle $\phi_s$ around a source ring $R_s=\bar{R}_s$ in the NHEK, wrapping around it multiple times as $\epsilon\to0$.

\begin{figure}
	\centering
	\includegraphics[width=\columnwidth]{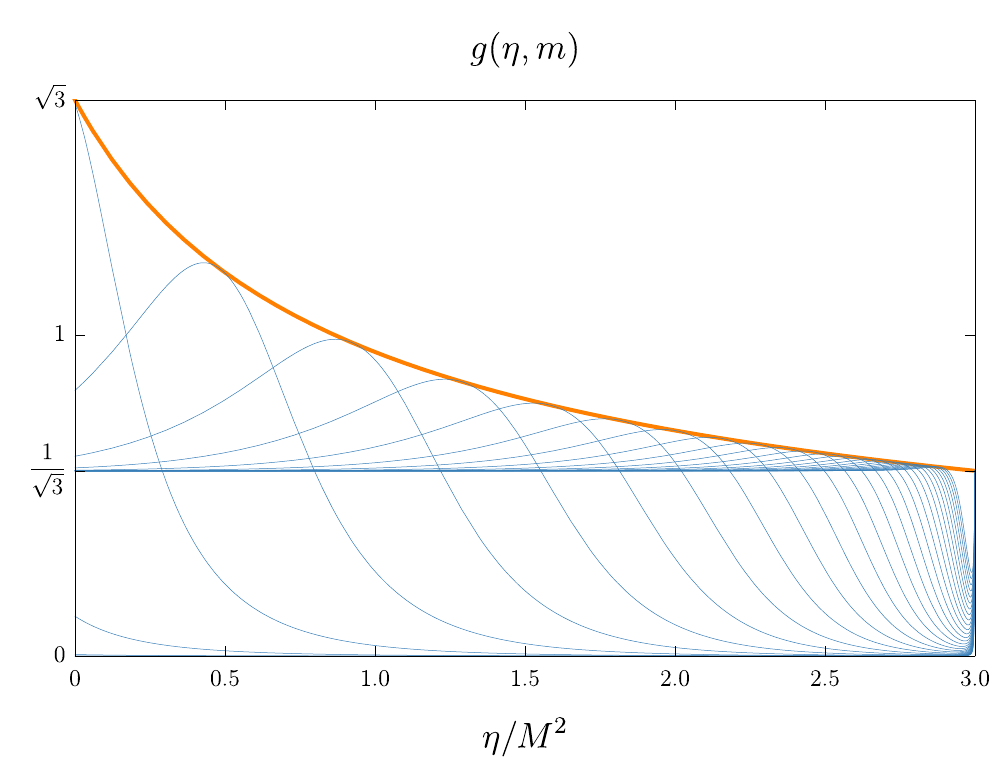}
	\caption{Observed redshift $g(\eta,m)$ of photons emitted from an equatorial orbiter on the ISCO of a near-extremal black hole with spin $a=M\sqrt{1-10^{-18}}$, shown for $m\in\br{0,30}$ (blue), and overlaid with the envelope $g_0(\eta)$ in Eq.~\eqref{eq:RedshiftEnvelope} (orange).}
	\label{fig:Envelope}
\end{figure}

In the NHEK regime, the observed redshift \eqref{eq:Redshift} takes the leading-order form (Eq.~(3.11) of Ref.~\cite{Gralla2017})
\begin{align}
	\label{eq:RedshiftNHEK}
	g=\pa{\sqrt{3}+\frac{4}{\sqrt{3}}\frac{\bar{\lambda}}{\bar{R}}}^{-1}.
\end{align}
Plugging in Eq.~\eqref{eq:GeodesicsNHEK} yields the redshift $g(\eta,m,\bar{R})$ of photons received with conserved quantity $\eta=M^2(3-q^2)$ from NHEK radius $\bar{R}$ after encountering $m$ turning points and sweeping through an azimuth $\Delta\phi(q)$ [Eq.~\eqref{eq:SourceAngle}] along their trajectory.  In Fig.~\ref{fig:Envelope}, we plot these curves over their full range $\eta\in\br{0,3M^2}$ and for $m\in\cu{0,1,2,\ldots,30}$ in the case of emission from the ISCO ($\bar{R}=2^{1/3}$) of a near-extremal BH with $\epsilon=10^{-6}$.

As reviewed in Sec.~\ref{sec:BlueBlade} above, the transfer function formalism developed by Cunningham to study emission from equatorial disks only considers their direct $\bar{m}=0$ emission.  But Fig.~\ref{fig:Envelope} shows that as one varies the elevation $\beta$ along the NHEKline, one must also vary the image label $m$ in order to pick out the maximally blueshifted photons.  It was empirically observed in Sec.~4.1 of Ref.~\cite{Gralla2017} that the choice of $m$ that maximizes $g$ is the one such that the emission radius is a radial turning point for the photon's geodesic. This condition requires a specific choice $m=m_0(q)$ given in Eq.~(4.3) of Ref.~\cite{Gralla2017}.  This choice is such that $m_0(q)\sim\log{\epsilon}$, which ensures that $\Upsilon_{m_0}(q)\sim\O{\epsilon^0}$ as required by consistency of the expansion, and leads us to define (Eq.~(4.5) of Ref.~\cite{Gralla2017})
\begin{align}
	\label{eq:Envelope}
	\bar{\lambda}_0=\bar{\lambda}(\eta(q),m_0(q))
	=\pa{\sqrt{1-\frac{q^2}{4}}-1}\bar{R}.
\end{align}
Plugging this into Eq.~\eqref{eq:RedshiftNHEK} defines a function
\begin{align}
	\label{eq:RedshiftEnvelope}
	g_0=\frac{\sqrt{3}}{2\sqrt{4-q^2}-1}
	=\frac{\sqrt{3}}{2\sqrt{1+\eta/M^2}-1},
\end{align}
which is plotted in orange in Fig.~\ref{fig:Envelope}, where it is overlaid on top of the functions $g(\eta,m)$ for which it provides the envelope.  Note that for every $m$, $\bar{\lambda}(\eta,m)$ in Eq.~\eqref{eq:GeodesicsNHEK} is an approximate solution at small $0<\epsilon\ll1$ of the geodesic equation \eqref{eq:NullGeodesics}.  On the other hand, $\bar{\lambda}_0$ is not a solution unless it coincides with $\bar{\lambda}(\eta,m)$ for some $m$, which happens precisely when the real function $m_0(q)$ is an exact integer.  Whenever this happens, the corresponding curve $g(\eta,m)$ coincides with the envelope $g_0(\eta)$.  It can be seen in Fig.~\ref{fig:Envelope} that these intersections are already frequent at $\epsilon=10^{-6}$, and they become more dense as $\epsilon\to0$.  Therefore, the envelope approximation $g_0(\eta)$ improves in the extremal limit.  This insight was used as the starting point of the analysis in Ref.~\cite{Lupsasca2018}, which extended Cunningham's formalism to the NHEK portion of a disk by using this envelope to take a ``diagonal cut'' across $m$.

For our purposes, it suffices to plug Eq.~\eqref{eq:NHEKlinePosition} into Eq.~\eqref{eq:RedshiftEnvelope} to obtain the high-spin, high-inclination approximation to the observed redshift,
\begin{align}
	g_0(\beta,x)\approx\frac{\sqrt{3}x}{2\sqrt{\pa{x^2-2}^2+\pa{\frac{x\beta}{M}}^2}-x}.
\end{align}
For any fixed inclination $x=\sin{\theta_o}$, $\pd_\beta g_0\le0$ for $\beta\ge0$, so the observed blueshift is monotonically decreasing in $\ab{\beta}$.  In particular, the MOB appears at the center $\beta=0$ of the NHEKline, where we find
\begin{align}
	\label{eq:HighSpinMOB}
	g_\mathrm{mob}(x)=g_0(0,x),
\end{align}
in agreement with Eq.~\eqref{eq:ExtremalEnvelope} above.

\acknowledgements{We acknowledge Laura Brenneman, Thomas Dauser, Samuel Gralla, Achilleas Porfyriadis, and Andrew Strominger for useful discussions.  DG acknowledges support from NSF GRFP grant DGE1144152.  SH and AL gratefully acknowledge support from the Jacob Goldfield Foundation.  AL also thanks Ori Goldfield and Abhishek Pathak for helpful comments.}

\appendix

\section{Transfer functions for equatorial disks}
\label{app:TransferFunctions}

\subsection{Radial roots}

In Eq.~\eqref{eq:EquatorialRadius}, we gave an analytic expression for the transfer functions $r_s^{(m)}(\alpha,\beta)$ in terms of the radial roots $\cu{r_1,r_2,r_3,r_4}$ of the radial potential $\mathcal{R}(r)$.  These roots were explicitly computed in Ref.~\cite{Gralla2020b} and their use in this context was described in App.~A of Ref.~\cite{Gralla2020a}.  Here, we summarize these results for the reader's convenience.

We introduce auxiliary quantities
\begin{subequations}
\label{eq:AuxiliaryQuantities}
\begin{align}
	\mathcal{A}&=a^2-\eta-\lambda^2,\\
	\mathcal{B}&=2M\br{\eta+\pa{\lambda-a}^2} >0,\\
	\mathcal{C}&=-a^2\eta,
\end{align}
\end{subequations}
in terms of which we define
\begin{subequations}
\begin{align}
	\mathcal{P}&=-\frac{\mathcal{A}^2}{12}-\mathcal{C},\\
	\mathcal{Q}&=-\frac{\mathcal{A}}{3}\br{\pa{\frac{\mathcal{A}}{6}}^2-\mathcal{C}}-\frac{\mathcal{B}^2}{8},\\
	\omega_\pm&=\sqrt[3]{-\frac{\mathcal{Q}}{2}\pm\sqrt{\pa{\frac{\mathcal{P}}{3}}^3+\pa{\frac{\mathcal{Q}}{2}}^2}},\\
	z&=\sqrt{\frac{\omega_++\omega_-}{2}-\frac{\mathcal{A}}{6}}
	>0.
\end{align}
\end{subequations}
The radial roots are then given by
\begin{subequations}
\label{eq:RadialRoots}
\begin{align}
	r_1&=-z-\sqrt{-\frac{\mathcal{A}}{2}-z^2+\frac{\mathcal{B}}{4z}}, \\
	r_2&=-z+\sqrt{-\frac{\mathcal{A}}{2}-z^2+\frac{\mathcal{B}}{4z}},\\
	r_3&=z-\sqrt{-\frac{\mathcal{A}}{2}-z^2-\frac{\mathcal{B}}{4z}},\\
	\label{eq:r4}
	r_4&=z+\sqrt{-\frac{\mathcal{A}}{2}-z^2-\frac{\mathcal{B}}{4z}}.
\end{align}
\end{subequations}
Together with the inverse of Eqs.~\eqref{eq:BardeenCoordinates},
\begin{align}
	\lambda=-\alpha\sin{\theta_o},\quad
	\eta=\pa{\alpha^2-a^2}\cos^2{\theta_o}+\beta^2,
\end{align}
these relations explicitly define the roots $r_i(\alpha,\beta)$, and hence the transfer functions $r_s^{(m)}(\alpha,\beta,\theta_o)$ describing the optical appearance of a thin equatorial disk surrounding a Kerr black hole.

\subsection{Polar (face-on) observers}

For an observer on the spin axis $\theta_o=0^\circ$, the Cartesian system \eqref{eq:BardeenCoordinates} degenerates.  One must instead use polar coordinates $(b,\varphi)$ with impact radius (Eq.~(61) of Ref.~\cite{Gralla2020a})
\begin{align}
	b=\sqrt{\eta+a^2},
\end{align}
in terms of which (since $\lambda=0$) Eqs.~\eqref{eq:AuxiliaryQuantities} simplify to
\begin{subequations}
\begin{align}
	\mathcal{A}&=2a^2-b^2,\\
	\mathcal{B}&=2Mb^2,\\
	\mathcal{C}&=a^2\pa{a^2-b^2}.
\end{align}
\end{subequations}
The integral $G_\theta$ also simplifies to (Eq.~(78) of Ref.~\cite{Gralla2020a})
\begin{align}
	G_\theta^{\bar{m}}&=\frac{2\bar{m}+1}{\sqrt{b^2-a^2}}K\pa{\frac{a^2}{a^2-b^2}}.
\end{align}
Hence, the images of any equatorial source ring $r=r_s$ are circles of constant $b$, with the exact relation obtained by plugging these formulas into Eq.~\eqref{eq:EquatorialRadius}.  This yields the transfer functions $r_s^{(\bar{m})}(b)$ that describe the optical appearance of an equatorial disk viewed by a polar observer.  Series expanding the $\bar{m}=0$ transfer function $r_s^{(0)}(b)$ in small parameter $0<M/b\ll1$ (or equivalently, in large source radius $r_s\gg M$) directly produces Eq.~\eqref{eq:PolarTransferFunction}.

To help with this expansion, we note that to $\O{\frac{M^2}{b^2}}$, the roots \eqref{eq:RadialRoots} are
\begin{subequations}
\begin{align}
	r_1&=-b-M+\frac{3M^2+a^2}{2b}-\frac{4M^3}{b^2},\\
	r_2&=M-\sqrt{M^2-a^2}+\frac{4M^3-\frac{4M^4-2a^2M^2}{\sqrt{M^2-a^2}}}{b^2},\\
	r_3&=M+\sqrt{M^2-a^2}+\frac{4M^3+\frac{4M^4-2a^2M^2}{\sqrt{M^2-a^2}}}{b^2},\\
	r_4&=b-M-\frac{3M^2+a^2}{2b}-\frac{4M^3}{b^2},
\end{align}
\end{subequations}
which implies that $k\sim\O{\frac{M}{b}}$.  To obtain Eq.~\eqref{eq:PolarTransferFunction}, one must plug these expressions for the roots, together with
\begin{align}
	G_\theta^{0}=\frac{\pi}{2b}+\frac{\pi a^2}{8b^3}+\O{\frac{M^5}{b^5}},
\end{align}
into Eq.~\eqref{eq:EquatorialRadius}, and then use the standard expansions of the incomplete elliptic integral of the first kind $F(z|k)$ and the Jacobi function $\sn(z,k)$ in small parameter $k$.

\section{Schwarzschild black holes}

The computation of the geodesic integrals $I_r$ and $G_\theta$ [Eq.~\eqref{eq:NullGeodesics}] (as well as $I_i$ and $G_i$ for $i\in\cu{t,\phi}$) was carried out in Ref.~\cite{Gralla2020b} for rotating black holes with nonzero spin $0<a<M$.  In this appendix, we present simplified expressions for these integrals in the case of a nonrotating ($a=0$) Schwarzschild black hole.  This also results in simplified transfer functions for equatorial disks.

\subsection{Angular geodesic integrals}

The angular path integrals $G_\theta$, $G_t$, and $G_\phi$ are given in Eq.~(79) of Ref.~\cite{Kapec2020}.  In particular, for photons emitted from equatorial sources with $\cos{\theta_s}=0$ and encountering $m$ turning points along their trajectory to the observer,
\begin{align}
	G_\theta^m=\frac{1}{\sqrt{\eta+\lambda^2}}\br{\pi m\mp_o\arcsin\pa{\sqrt{1+\frac{\lambda^2}{\eta}}\cos{\theta_o}}}.
\end{align}
This formula may also be obtained directly from Eq.~\eqref{eq:AngularMinoTime} by expanding to leading order in $a$.

In terms of polar coordinates $(b,\varphi)$ defined on the observer screen by $\alpha=b\cos{\varphi}$ and $\beta=b\sin{\varphi}$, we have
\begin{align}
	\eta+\lambda^2=b^2,\quad
	1+\frac{\lambda^2}{\eta}=\frac{1}{\cos^2{\theta_o}\cos^2{\varphi}+\sin^2{\varphi}}.
\end{align}
Smooth contours on the observer screen are labeled not by $m$ but rather by $\bar{m}=m-H(\beta)$, where $H$ denotes the Heaviside function (see Eq.~(82) of Ref.~\cite{Gralla2020a}).  Plugging these relations into $G_\theta^{\bar{m}}$ simplifies the geodesic integral to
\begin{align}
	\label{eq:SchwarzschildAngularIntegral}
	G_\theta^{\bar{m}}=\frac{1}{b}\br{\pa{\bar{m}+\frac{1}{2}}\pi+\arctan\pa{\sin{\phi}\tan{\theta_o}}}.
\end{align}

\subsection{Radial geodesic integrals}

We now analyze the radial path integrals using the methods of Ref.~\cite{Gralla2020b}.  The radial potential is
\begin{align}
	\mathcal{R}(r)=r^4-r\pa{r-2M}b^2.
\end{align}
By Cardano's method, the four roots of this quartic polynomial are
\begin{subequations}
\label{eq:SchwarzschildRoots}
\begin{align}
	r_1&=-\frac{2b}{\sqrt{3}}\cos\br{\frac{1}{3}\arccos\pa{\frac{b_c}{b}}},\\
	r_2&=0,\\
	r_3&=\frac{2b}{\sqrt{3}}\sin\br{\frac{1}{3}\arcsin\pa{\frac{b_c}{b}}},\\
	r_4&=\frac{2b}{\sqrt{3}}\cos\br{\frac{1}{3}\arccos\pa{-\frac{b_c}{b}}},
\end{align}
\end{subequations}
where $b_c=3\sqrt{3}M$ is the apparent radius of the critical curve in the axisymmetric case $\theta_o=0$.  These expressions are equivalent to Eqs.~\eqref{eq:RadialRoots} with $a=0$.  In general Kerr, these roots can exhibit four different behaviors, labeled I, II, III and IV in Ref.~\cite{Gralla2020b}. In Schwarzschild, however, only two of these behaviors are allowed:
\begin{enumerate}
	\item[I.] Four real roots, two outside horizon:\\
	$r_1<r_2<r_-<r_+<r_3<r_4$.
	\item[III.] Two real roots, both inside horizon:\\
	$r_1<r_2<r_-<r_+$ and $r_3=\bar{r}_4$.
\end{enumerate}
Only type I and type III light rays can reach a distant observer, whereupon they appear outside and inside the critical curve on the observer screen, respectively.  The relevant expressions for the integrals $I_i$ with $i\in\cu{t,r,\phi}$ are respectively given by cases (2) and (3) of Ref.~\cite{Gralla2020b}.

\subsection{Transfer functions for equatorial disks}

The inversion of $I_r=G_\theta$ for light rays shot back from the observer screen outside the critical curve $b=b_c$ is given by Eq.~\eqref{eq:EquatorialRadius} with $I_r=G_\theta^{\bar{m}}$ as given in Eq.~\eqref{eq:SchwarzschildAngularIntegral} and the radial roots $\cu{r_1,r_2,r_3,r_4}$ as given in Eq.~\eqref{eq:SchwarzschildRoots}.  This fully specifies the transfer functions $r_s^{\bar{m}}(b,\varphi,\theta_o)$ describing the optical appearance of equatorial disks around a Schwarzschild black hole.

Since the equatorial disk breaks the rotational symmetry of the underlying spacetime, its image is axisymmetric only for polar observers: only when $\theta_o=0^\circ$ does the dependence on $\varphi$ drop out from Eq.~\eqref{eq:SchwarzschildAngularIntegral}.

The formulas for the transfer functions $r_s^{\bar{m}}(b,\varphi,\theta_o)$ (as well as the radial integrals $I_r$, $I_t$, and $I_\phi$) can be greatly simplified using the relations
\begin{subequations}
\begin{align}
	r_{31}&=2b\sin\br{\frac{1}{3}\arccos\pa{-\frac{b_c}{b}}},\\
	r_{41}&=2b\cos\br{\frac{1}{3}\arcsin\pa{\frac{b_c}{b}}},\\
	k&=\frac{\sin\br{\frac{2}{3}\arcsin\pa{\frac{b_c}{b}}}}{\sin\br{\frac{2}{3}\arccos\pa{-\frac{b_c}{b}}}},\\
	\frac{1}{2}\sqrt{r_{31}r_{42}}&=b\sqrt{\frac{\sin\br{\frac{2}{3}\arccos\pa{-\frac{b_c}{b}}}}{2\sqrt{3}}}.
\end{align}
\end{subequations}

\section{Near-extremal black holes}

The results of Sec.~\ref{subsec:HighInclinationHighSpin} rely on the asymptotic expansion as $\epsilon\to0$ of the radial integral $I_r$, which was first computed (along with that of $I_\phi$ and $I_t$) in Refs.~\cite{Porfyriadis2017,Gralla2017}.  In this appendix, we describe how these asymptotic forms may be recovered as a special case of the general formulas recently derived in Ref.~\cite{Gralla2020b} by taking the near-extremal limit $\epsilon\to0$.  This requires the use of a nontrivial identity, whose derivation we now turn to.

\subsection{Asymptotic expansion of the complete elliptic integral of the first kind}

In this section, we prove that the incomplete elliptic integral of the first kind $F(z|k)$, which we define as
\begin{align}
	\label{eq:EllipticF}
	F(z|k)=\int_0^z\frac{\ed\theta}{\sqrt{1-k\sin^2{\theta}}},
\end{align}
admits the asymptotic expansion for small $0<\epsilon\ll1$,
\begin{subequations}
\label{eq:AsymptoticExpansion}
\begin{gather}
	F\pa{\left.\frac{\pi}{2}-\epsilon\right|1-A\epsilon^2}\stackrel{\epsilon\to0}{\approx}-\frac{1}{2}\log\pa{A\epsilon^2}+C,\\
	C=\log{4}-\log\pa{\frac{1}{\sqrt{A}}+\sqrt{1+\frac{1}{A}}},
\end{gather}
\end{subequations}
up to $\O{\epsilon^2\log{\epsilon}}$ corrections.

To prove this relation, we invoke Eq.~(17.4.13) of Ref.~\cite{Abramowitz1972}, which states that whenever $k=\sin^2{\alpha}$ and
\begin{align}
	\label{eq:Psi}
	\cos{\alpha}\tan{\phi}\tan{\psi}=1,
\end{align}
the following relation holds:
\begin{align}
	\label{eq:MagicIdentity}
	F(\phi|k)+F(\psi|k)=K(k).
\end{align}
As usual, $K(k)=F(\pi/2|k)$ denotes the complete elliptic integral of the first kind.  Since we are interested in  
\begin{align}
	\lim_{\epsilon\to0}F\pa{\left.\frac{\pi}{2}-\epsilon\right|1-A\epsilon^2},
\end{align}
where $A>0$, we set $\phi=\pi/2-\epsilon$ and $\cos^2{\alpha}=A\epsilon^2$, in which case Eq.~\eqref{eq:Psi} gives
\begin{align}
	\psi_A=\arctan\pa{\frac{1}{\sqrt{A}}}.
\end{align}
Plugging this into Eq.~\eqref{eq:MagicIdentity} then yields
\begin{align}
	\label{eq:IntermediateStep}
	F\pa{\left.\frac{\pi}{2}-\epsilon\right|1-A\epsilon^2}\stackrel{\epsilon\to0}{\approx}&K\pa{1-A\epsilon^2}\\
	&-F\pa{\left.\arctan\pa{\frac{1}{\sqrt{A}}}\right|1},\nonumber
\end{align}
where we set the parameter of the last term to 1, neglecting corrections.  Integrating Eq.~\eqref{eq:EllipticF} at $k=1$ results in
\begin{align}
	F(z|1)=\log\pa{\frac{2}{1-\tan\pa{\frac{z}{2}}}-1},\quad
	0\le z<\frac{\pi}{2}.
\end{align}
In particular, this implies that
\begin{align}
	F\pa{\left.\arctan\pa{\frac{1}{\sqrt{A}}}\right|1}=\log\pa{\frac{1}{\sqrt{A}}+\sqrt{1+\frac{1}{A}}}.
\end{align}
Next, we can use the asymptotic expansion of the complete elliptic integral,
\begin{align}
	K\pa{1-A\epsilon^2}\stackrel{\epsilon\to0}{\approx}-\frac{1}{2}\log\pa{A\epsilon^2}+\log{4},
\end{align}
up to $\mathcal{O}(\epsilon^2\log\epsilon)$ corrections.  Plugging these expressions into Eq.~\eqref{eq:IntermediateStep} results in the identity \eqref{eq:AsymptoticExpansion}.

\subsection{Asymptotic expansion of geodesic integrals}

In the near-extremal regime of Ref.~\cite{Gralla2017}, where
\begin{subequations}
\label{eq:NearExtremalParameters}
\begin{align}
	a&=M\sqrt{1-\epsilon^3},\\
	r&=M\pa{1+R},\quad
	R=\epsilon\bar{R}\\
	\lambda&=2M\pa{1-\epsilon\bar{\lambda}},\\
	\eta&=M^2\pa{3-q^2},
\end{align}
\end{subequations}
the radial roots $r_i=M\pa{1+R_i}$ in Eqs.~\eqref{eq:RadialRoots} take the following form at leading order in $\epsilon$ (assuming $\bar{\lambda}\ge0$):
\begin{subequations}
\label{eq:NearExtremalRoots}
\begin{align}
	R_1&=-2-\sqrt{4-q^2},
	&&R_3=\frac{2\bar{\lambda}R_1}{q^2}\epsilon,\\
	R_2&=-2+\sqrt{4-q^2},
	&&R_4=\frac{2\bar{\lambda}R_2}{q^2}\epsilon.
\end{align}
\end{subequations}
In particular, $R_1\sim R_2\sim\O{\epsilon}$, while $R_3\sim R_4\sim\O{\epsilon^0}$.

We want to compute the radial integral $I_r$ appearing in Eq.~\eqref{eq:NullGeodesics} in the near-extremal regime \eqref{eq:NearExtremalParameters} with $R_o$ remaining finite in $\epsilon$ but $R_s=\epsilon\bar{R}_s$ scaling into the NHEK.  In this regime, $I_r$ falls under case (2) of Ref.~\cite{Gralla2020b}, where it is given in Eq.~(B40) as (with $k$ given in Eq.~\eqref{eq:k} above)
\begin{align}
	I_r&=I_r^\mathrm{far}-I_r^\mathrm{near},\\
	I_r^\mathrm{far}&=\frac{2}{\sqrt{r_{31}r_{42}}}F(z(R_o)|k),\\
	I_r^\mathrm{near}&=\frac{2}{\sqrt{r_{31}r_{42}}}F\pa{\left.z\pa{\epsilon\bar{R}_s}\right|k},\\
	z(R)&=\arcsin\sqrt{\frac{R-R_4}{R-R_3}\frac{r_{31}}{r_{41}}}.
\end{align}
Plugging in the near-extremal regime parameters \eqref{eq:NearExtremalParameters} and roots \eqref{eq:NearExtremalRoots} into these expressions and expanding to leading order in $\epsilon$ results in
\begin{align}
	I_r^\mathrm{far}&=-\frac{1}{q}\log{\epsilon}+\frac{1}{q}\log\br{\frac{2q^4R_o}{\sqrt{4-q^2}\pa{q^2+qD_o+2R_o}\bar{\lambda}}},\nonumber\\
	I_r^\mathrm{near}&=-\frac{1}{q}\log{\frac{2\sqrt{4-q^2}\bar{\lambda}}{q^2\bar{R}_s+qD_s+4\bar{\lambda}}},
\end{align}
with $D_o^2=q^2+4R_o+R_o^2$ and $D_s^2=q^2\bar{R}_s^2+8\bar{\lambda}\bar{R}_s+4\bar{\lambda}^2$.

Combining these two terms reproduces the asymptotic form of $I_r$ first derived using the method of matched asymptotic expansions in Refs.~\cite{Porfyriadis2017,Gralla2017}.  The expansion of $I_r^\mathrm{far}$ crucially requires the use of the asymptotic formula \eqref{eq:AsymptoticExpansion}.

\bibliographystyle{utphys}
\bibliography{MaxBlueshift.bib}

\end{document}